\documentclass[structabstract]{aa}
\newcommand{\Mea}{\hbox{\hbox{M}$_{\oplus}\;$}}

\newcommand{\Meao}{\hbox{\hbox{M}$_{\oplus}$}}

\newcommand{\Msun}{\hbox{\hbox{M}$_\odot\;$}}

\newcommand{\kmso}{\hbox{${\rm km}\:{\rm s}^{-1}$}}
\newcommand{\teff}{$T_{\rm eff}\;$}

\newcommand{\loggl}{$\log (g/{\rm cm~s}^2)$}

\usepackage{longtable}
\usepackage{lscape}
\usepackage{natbib}
\bibpunct{(}{)}{;}{a}{}{,} 
\usepackage{graphicx}
\usepackage{txfonts}
\begin{document}
\title{Searching for the signatures of terrestrial planets in F-,
G-type main-sequence stars
\thanks{Based on observations collected with the HARPS spectrograph 
at the 3.6-m telescope (072.C-0488(E)), installed at the 
La Silla Observatory, ESO (Chile), with the UVES 
spectrograph at the 8-m Very Large Telescope (VLT) 
(program IDs: 67.C-0206(A), 074.C-0134(A), 075.D-0453(A)), 
installed at the Cerro Paranal Observatory, ESO (Chile), 
and with the UES spectrograph at the 4.2-m William Herschel 
Telescope (WHT), installed at the Spanish Observatorio del Roque 
de los Muchachos of the Instituto de Astrof{\'\i}sica de Canarias, 
on the island of La Palma}
}

\titlerunning{Signatures of terrestrial planets in F-, G- analogs}
\authorrunning{Gonz\'alez Hern\'andez et al.}


\author{
J.~I.~Gonz\'alez~Hern\'andez\inst{1,2}
\and E.~Delgado-Mena\inst{3} \and S.~G.~Sousa\inst{3} \and 
G.~Israelian\inst{1,2} \and N.~C.~Santos\inst{3,4} \and
V.~Zh.~Adibekyan\inst{3} \and S.~Udry\inst{5}	  
}


\institute{
Instituto de Astrof{\'\i }sica de Canarias (IAC), 
E-38205 La Laguna, Tenerife, Spain\\
\email{jonay@iac.es}
\and
Depto. Astrof{\'\i }sica, Universidad de La 
Laguna (ULL), E-38206 La Laguna, Tenerife, Spain
\and
Centro de Astrof\'isica, Universidade do Porto, Rua
das Estrelas, 4150-762 Porto, Portugal 
\and
Departamento de F\'{\i}sica e Astronomia, Faculdade 
de Ci\^encias, Universidade do Porto, Portugal
\and
Observatoire Astronomique de l'Universit\'e de
Gen\`eve, 51 Ch. des Maillettes, -Sauverny- Ch1290, Versoix, 
Switzerland
}

\date{Received July 24, 2012; accepted January 9, 2013}

 
\abstract
{Detailed chemical abundances of volatile and refractory elements 
have been discussed in the context of terrestrial-planet formation 
during in past years.}
{The HARPS-GTO high-precision planet-search program has provided an
extensive database of stellar spectra, which we have inspected in 
order to select the best-quality spectra available for late type 
stars. We study the volatile-to-refractory abundance 
ratios to investigate their possible relation with the low-mass
planetary formation.}
{We present a fully differential chemical abundance analysis using 
high-quality HARPS and UVES spectra of 61 late F- and early G-type 
main-sequence stars, where 29 are planet hosts and 32 are stars 
without detected planets.}
{As for the previous sample of solar analogs, these stars slightly 
hotter than the Sun also provide very accurate Galactic chemical 
abundance trends in the metallicity range $-0.3<{\rm [Fe/H]}<0.4$.
Stars with and without planets show similar mean abundance ratios. 
Moreover, when removing the Galactic chemical evolution
effects, these mean abundance ratios, 
$\Delta {\rm [X/Fe]_{SUN-STARS}}$, against condensation temperature 
tend to exhibit less steep trends with nearly zero or slightly 
negative slopes.
We have also analyzed a subsample of 26 metal-rich stars,
13 with and 13 without known planets, with spectra at S/N~$\sim 850$, 
on average, in the narrow metallicity range $0.04<{\rm [Fe/H]}<0.19$. 
We find the similar, although not equal, abundance pattern with
negative slopes for both samples of stars with and without planets. 
Using stars at S/N~$\ge 550$ provides equally steep abundance trends 
with negative slopes for stars both with and without planets.
We revisit the sample of solar analogs to study the abundance patterns  
of these stars, in particular, 8 stars hosting super-Earth-like
planets. Among these stars having very low-mass planets, only four of
them reveal clear increasing abundance trends versus condensation
temperature.
 }
{Finally, we compared these observed slopes with those predicted
using a simple model that enables us to compute the mass of rocks 
that have formed terrestrial planets in each planetary system.
We do not find any evidence supporting the conclusion that the 
volatile-to-refractory abundance ratio is related to the presence 
of rocky planets.}


\maketitle

\section{Introduction\label{secintro}}

In the past decade, the exhaustive search for extrasolar planets 
has given rise to a significant collection of high-quality 
spectroscopic data \citep[see e.g.][]{mar05,udr07}. 
In particular, the HARPS-GTO high-precision planet-search program
\citep{may03} 
has conducted in the first five years an intensive tracking of F-, 
G-, and K-type main-sequence stars in the solar neighborhood, 
resulting in the discovery of an increasing population of Neptune 
and super-Earth-like planets \citep{mau08,may11}. 
Interestingly, the very strong correlation
observed between the host star metallicity \citep{san01,san04,vaf05} 
and the frequency of giant planets appears to almost 
vanish when dealing with these less massive planets
\citep{udr06,sou08,sou11,may11,adi12b,buc12}. 
In addition, most of Neptune and super-Earth-like planets are 
found in multiplanetary systems \citep[see e.g.][]{may09}.
\citet{may11} have recently announced the discovery of 41 planets
with HARPS, among which there are 16 super-Earth-like planets. 
Many of these planets belong to complex planetary systems with
a variety of planet masses and orbital properties 
\citep[see e.g.][]{lov11}, thus challenging our understanding 
of planet formation and evolution.

The search for peculiarities in the chemical abundance patterns of
planet-host stars has also been done 
\citep[see e.g.][]{gil06,ecu06,nev09,adi12a}, it suggests that there
are only small differences in some element abundance ratios 
between ``single'' stars 
(hereafter, ``single'' refers to stars without detected planets) 
and planet hosts 
\citep[see the introduction in][for a more detailed background]{gon10}.
However, based on HARPS and Kepler data, \citet{adi12c,adi12a} have 
found that planet-host stars show significant 
enhancement of alpha elements at metallicities slightly below 
solar and that they belong to the Galactic thick disk. 
They find that at metallicities [Fe/H]~$ < -0.3$~dex, 13 
planet hosts (HARPS and Kepler samples) out of 17 (about 76 \%) 
are enhanced in alpha elements, while only 32 \% of the stars without
known planets are alpha-enhanced at a given metallicity. 
Thus, planet formation requires a certain amount of metals, and
therefore, at the relatively low metallicity of[Fe/H]~$ < -0.3$~dex, 
planet-host stars tend to be alpha-enhanced and tend to belong 
to the chemically defined thick disk 
where the stars should have higher overall metallicity
\citep{adi12c}.
This result may be supported by theoretical studies based on 
core-accretion models \citep[see e.g.][]{mor12}.

Recently, \citet{mel09} have found an anomalous but small 
($\sim 0.1$~dex) volatile-to-refractory abundance ratio of a sample 
of 11 solar twins when compared to the solar ratio. They
interpret this particularity as a signature of
terrestrial planets in the solar planetary system. Their study
used high-S/N and high-resolution spectroscopic data. 
Later on, this
conclusion was also supported by \citet{ram09}, who analyzed a sample
of 64 solar analogs, but their result was not so 
evident, at least, not at the level of the accuracy of the 
abundance pattern shown in \citet{mel09}, where the element 
abundances are at a distance of $\leq 0.015$~dex from the mean 
abundance pattern~\citep[see also][]{ram10,sch11b}.  
\citet{gon10} analyzed a sample of seven solar twins, five without 
and two with planets, using very high-quality HARPS spectroscopic 
data, but maybe still not enough to confirm the result by 
\citet{mel09}.
However, \citet{gon10} find that on average both stars with and 
without detected planets seem to exhibit a very similar abundance 
pattern, even when considering a metal-rich subsample of 
14 planet hosts and 14 solar analogs. 
It is important to note that \citet{gon10} analyzed
stars with planets of different minimum masses from Jupiter-mass to
super-Earth planets. Nevertheless, the sample in \citet{ram09} 
contains same planet-host stars, but these authors did not 
distinguish and/or mention them, which may lead to confusing results. 
The results found by \citet{mel09} have not been fully
confirmed~\citep[see also][]{ram10,gon11sea,gui10,gui11,sch11a}, 
in particular the relation with the presence of planets is strongly
debated \citep{gon10,gon11iau}. Recently, they have however  
found a similar abundance pattern in the solar twin HIP~56948, which
they claim may have terrestrial planets~\citep{mel12}.
 
\begin{table*}
\caption{Spectroscopic observations}
\centering          
\begin{tabular}{lrrrrrrrr}
\noalign{\smallskip}
\noalign{\smallskip}
\noalign{\smallskip}
\hline\hline       
\noalign{\smallskip}
Spectrograph & Telescope & Spectral range &
$\lambda/\delta\lambda$ & Binning & $N_{\rm stars}$$^{\rm a}$ &
$\mid {\rm S/N} \mid$$^{\rm b}$ & 
$\delta \mid {\rm S/N} \mid$$^{\rm c}$ & 
$\Delta {\rm S/N} $$^{\rm d}$ \\
\noalign{\smallskip}
\hline       
\noalign{\smallskip}
 & & [{\AA}] & & [{\AA}/pixel] & & & & \\ 
\noalign{\smallskip}
\hline       
\noalign{\smallskip}
HARPS & 3.6m-ESO & 3800--6900 & 110,000 & 0.010 & 50 & 877 & 429 & 265--2015 \\ 
UVES  & 8m-VLT   & 4800--6800 &  85,000 & 0.016 & 7  & 722 & 150 & 510--940 \\ 
UES   & 4.2m-WHT  & 4150--7950 &  65,000 & 0.029 & 4  & 643 & 180 & 520--850 \\ 
\noalign{\smallskip}
\hline                  
\noalign{\smallskip}
\noalign{\smallskip}
\end{tabular}

\tablefoot{Details of the spectroscopic data used in this work.
The UVES and UES stars are all planet hosts. 
Three of the UVES stars were observed with a slit of $0.3\arcsec$
providing a resolving power $\lambda/\delta\lambda\sim 110,000$.
\tablefoottext{a}{Total number of stars including those with and
without known planets.} 
\tablefoottext{b}{Mean signal-to-noise ratio, at $\lambda = 
6070$~{\AA}, of the spectroscopic data used in this work.} 
\tablefoottext{c}{Standard deviation from the mean signal-to-noise 
ratio.} 
\tablefoottext{d}{Signal-to-noise range of the spectroscopic data.} 
}
\label{tblobs}      
\end{table*}

\begin{figure}[!ht]
\centering
\includegraphics[width=6.7cm,angle=90]{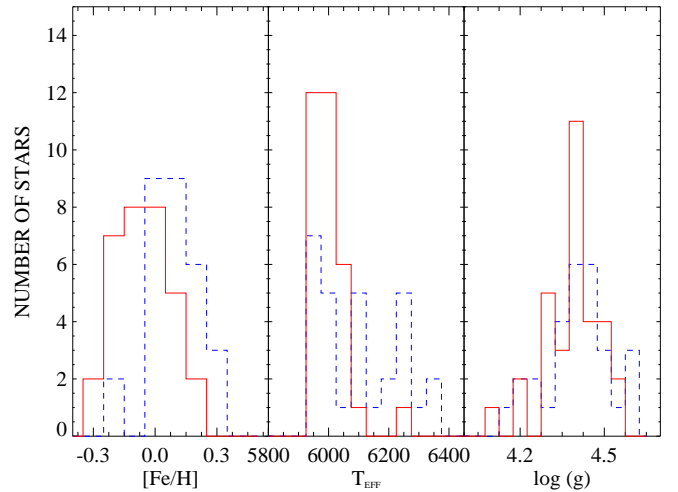}
\caption{ 
Histograms of the stellar parameters and metallicities 
of the whole sample of ``hot'' analogs hosting planets 
(blue dashed lines) and without known planets (red solid lines).
}
\label{fpar}
\end{figure}

\begin{figure*}[ht]
\centering
\includegraphics[width=11.5cm,angle=90]{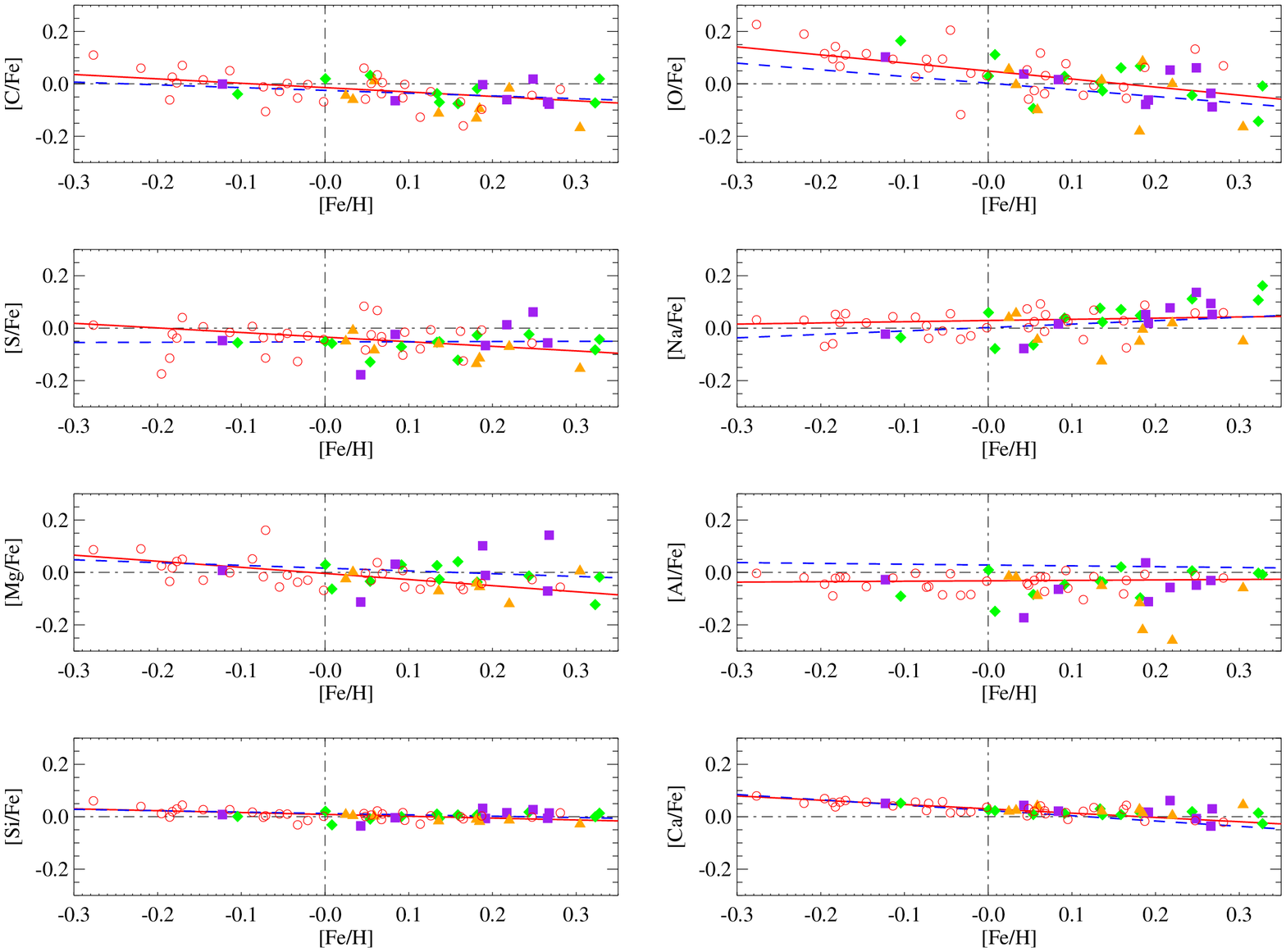}
\includegraphics[width=11.5cm,angle=90]{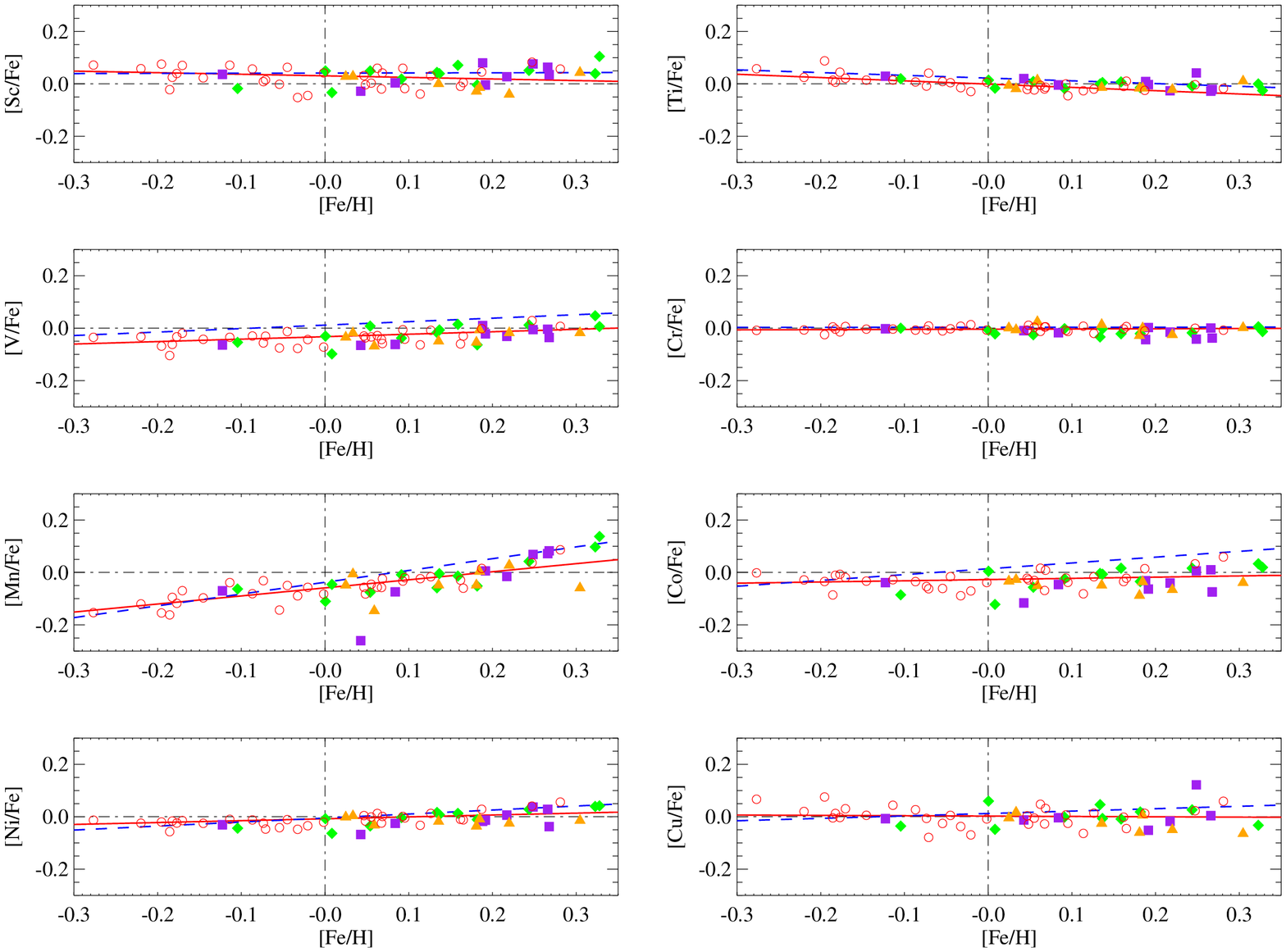}
\caption{ 
Chemical abundance ratios [X/Fe] versus [Fe/H] for the 
whole sample of ``hot'' analogs, containing 32 ``single'' stars 
(red open circles) and 29 stars hosting planets, with the most 
massive planet in an orbital period $P_{\rm orb} < 65$ 
days (green filled diamonds), $115 < P_{\rm orb} < 630$ days (violet 
filled squares), $950 < P_{\rm orb} < 4050$ days (orange filled triangles).  
Red solid lines provide linear fits to the data points of the 
32 ``single'' stars, whereas blue dashed lines show the linear fits 
to the data points of ``single'' solar analogs in \citet{gon10}.
}  
\label{fgala}
\end{figure*}

\begin{table}[!ht]
\centering
\caption[]{Ranges of stellar parameters and metallicities in 
different samples\label{tblsam}}
\begin{tabular}{lrrrr}
\noalign{\smallskip}
\noalign{\smallskip}
\noalign{\smallskip}
\hline
\hline
\noalign{\smallskip}
Sample$^{\rm a}$ & $T_{\mathrm{eff}}$ & $\log g$ & $\mathrm{[Fe/H]}$ 
& $N_{\rm stars}$$^{\rm b}$ \\
\noalign{\smallskip}
\hline
\noalign{\smallskip}
  & [K] & [dex] & [dex] &  \\
\noalign{\smallskip}
\hline
\noalign{\smallskip}
HA   & 5950--6400 & 4.14--4.60 & $-0.30$ -- $+0.35$ & 61 \\
mrHA & 5950--6400 & 4.14--4.60 & $+0.04$ -- $+0.19$ & 26 \\
\noalign{\smallskip}
\hline	     
\noalign{\smallskip}
\noalign{\smallskip}
\end{tabular}
\tablefoot{
\tablefoottext{a}{Stellar samples: ``hot'' analogs, ``HA'', 
metal-rich  ``hot'' analogs,  ``mrHA''.}
\tablefoottext{b}{Total number of stars including those with and
without known planets.}
}
\end{table}

\citet{del10} have shown that the chemical abundance pattern of the
planet-host stars can be used to distinguish among the 
different compositions of the possible terrestrial planets. 
In fact, theoretical simulations based on the observed variations in
the chemical abundances of host stars predict diverse compositions 
for extrasolar terrestrial planets, which are indeed quite different 
from those of the solar planetary system \citep{bon12}.  
 
In this paper, we investigate in detail the chemical abundances of 
``hot'' analogs (hereafter, ``hot'' analogs refers to late F- and
early G-type main-sequence stars hotter 
than the Sun, at temperatures $T_{\rm eff} \sim 6175$~K) 
of the HARPS GTO sample, which have 
narrower convective zones than Sun-like stars, with the aim of
searching for the enhanced signature of terrestrial planets in their
abundance patterns. We also reanalyzed the data of the solar analogs
presented in \citet{gon10} taking the new discovered
planets into account \citep[see e.g.][]{may11}. 
We pay special attention
to eight solar analogs harboring super-Earth-like planets under 
the assumption that these small planets may have similar 
chemical composition to that of terrestial or Earth-like planets.

\begin{table*}
\caption{Stellar parameters and metallicities from the UVES and UES
spectra}
\centering          
\begin{tabular}{lrrrrrrrr}
\noalign{\smallskip}
\noalign{\smallskip}
\noalign{\smallskip}
\hline\hline       
\noalign{\smallskip}
HD & $T_{\mathrm{eff}}$ & $\delta T_{\mathrm{eff}}$ & $\log
g$ & $\delta\log g$ & $\xi_{\rm t}$ & $\delta\xi_{\rm t}$ & $\mathrm{[Fe/H]}$&
$\delta\mathrm{[Fe/H]}$ \\
\noalign{\smallskip}
\hline                    
\noalign{\smallskip}
 & [K] & [K] & [dex] & [dex] & [\kmso] & [\kmso] & [dex] & [dex]\\  
\noalign{\smallskip}
\hline                    
\noalign{\smallskip}
209458$^{\rm a}$ & 6118 & 22 & 4.50 & 0.04 & 1.21 & 0.03 & 0.03  & 0.02\\
2039$^{\rm a}$   & 5959 & 31 & 4.45 & 0.04 & 1.26 & 0.03 & 0.32  & 0.02\\
213240       & 6022 & 19 & 4.27 & 0.05 & 1.25 & 0.02 & 0.14  & 0.01\\
23596$^{\rm b}$  & 6134 & 35 & 4.25 & 0.08 & 1.30 & 0.04 & 0.31  & 0.03\\
33636        & 5994 & 17 & 4.71 & 0.02 & 1.79 & 0.02 & -0.08 & 0.01\\
50554$^{\rm b}$  & 6129 & 50 & 4.41 & 0.07 & 1.11 & 0.06 & 0.01  & 0.04\\
52265$^{\rm a}$  & 6136 & 46 & 4.36 & 0.04 & 1.32 & 0.06 & 0.21  & 0.03\\
72659        & 5951 & 13 & 4.30 & 0.02 & 1.42 & 0.01 & 0.03  & 0.01\\
74156        & 6099 & 19 & 4.34 & 0.03 & 1.38 & 0.02 & 0.16  & 0.01\\
89744$^{\rm b}$  & 6349 & 43 & 3.98 & 0.05 & 1.62 & 0.05 & 0.22  & 0.03\\
9826 $^{\rm b}$  & 6292 & 43 & 4.26 & 0.06 & 1.69 & 0.05 & 0.13  & 0.03\\
\noalign{\smallskip}
\hline                  
\noalign{\smallskip}
\noalign{\smallskip}
\end{tabular}

\tablefoot{Stellar parameters, \teff and \loggl, microturbulent
velocities, $\xi_{\rm t}$, and metallicities , [Fe/H], and their
uncertainties, of the planet-host stars observed with UVES/VLT and 
UES/WHT spectrographs.\\
\tablefoottext{a}{Stars observed with the UVES spectrograph at VLT 
telescope at $\lambda/\delta\lambda\sim110,000$. The other stars were
observed with UVES at $\lambda/\delta\lambda\sim85,000$.} 
\tablefoottext{b}{Stars observed with the UES spectrograph at WHT 
telescope at $\lambda/\delta\lambda\sim65,000$.} 
}
\label{tblpar}      
\end{table*}

\begin{figure*}[ht]
\centering
\includegraphics[width=11.5cm,angle=90]{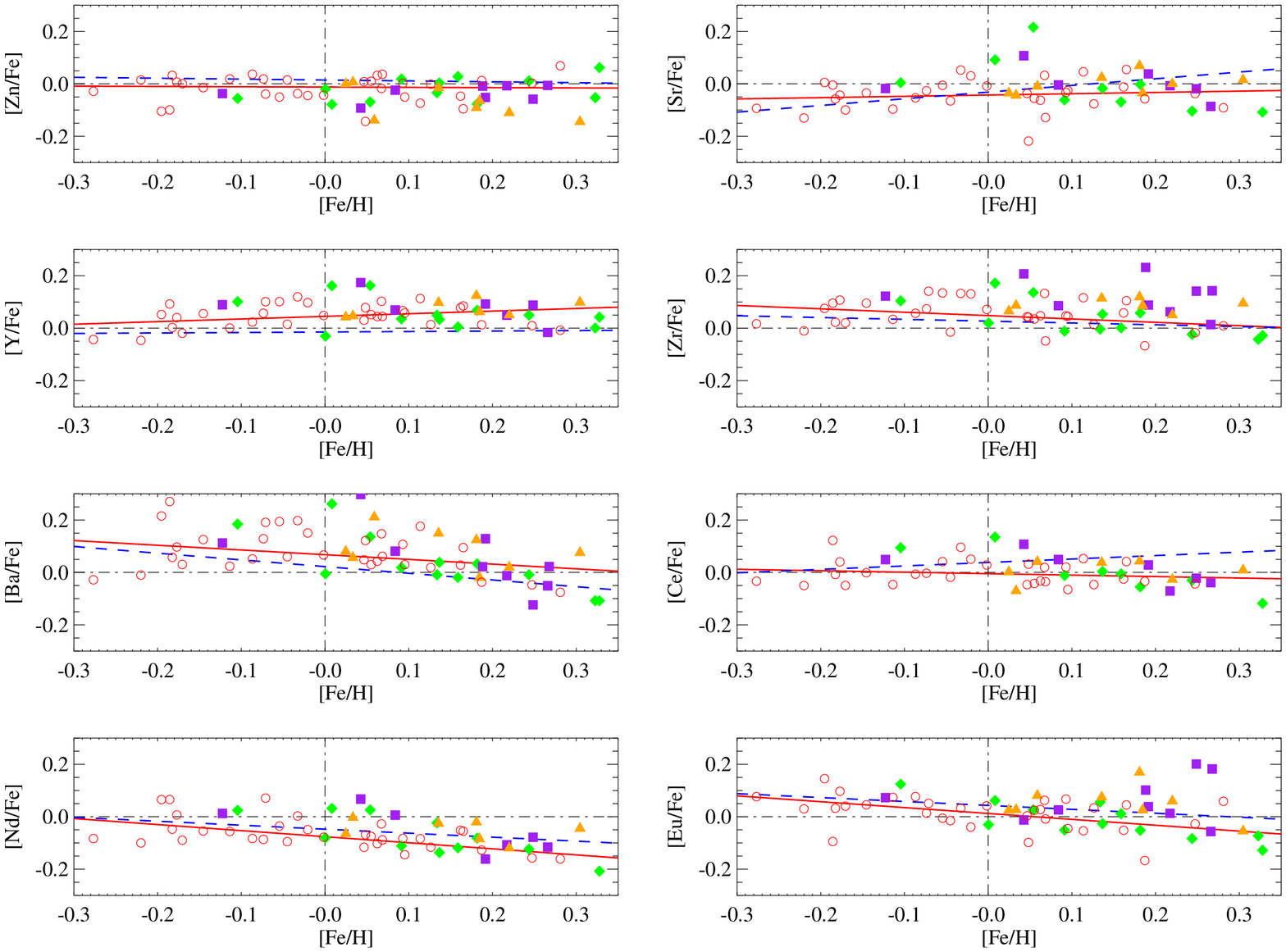}
\caption{ 
Fig.~\ref{fgala} continued.
}  
\label{fgalc}
\end{figure*}

\section{Observations\label{secobs}}

In this study, as in \citet{gon10}, we analyze
high-resolution and high-S/N spectroscopic data obtained 
with three different telescopes and instruments:
the 3.6m-ESO/HARPS at the {\itshape Observatorio
de La Silla} in Chile, the 8.2m-VLT/UVES at the 
{\itshape Observatorio Cerro Paranal} in Chile, and 
the 4.2m-WHT/UES at the {\itshape 
Observatorio del Roque de los Muchachos} in La Palma, Spain.  
In Table~\ref{tblobs}, we provide the wavelength ranges, resolving
power, S/N, and other details of these spectroscopic data. 
The data were reduced in a standard manner and later normalized
within the package IRAF\footnote{IRAF is distributed by 
the National Optical Observatory, which is operated by the 
Association of Universities for Research in Astronomy, Inc., 
under contract with the National Science Foundation.}, 
using low-order polynomial fits to the observed continuum. 

\begin{figure*}[!ht]
\centering
\includegraphics[width=6.7cm,angle=90]{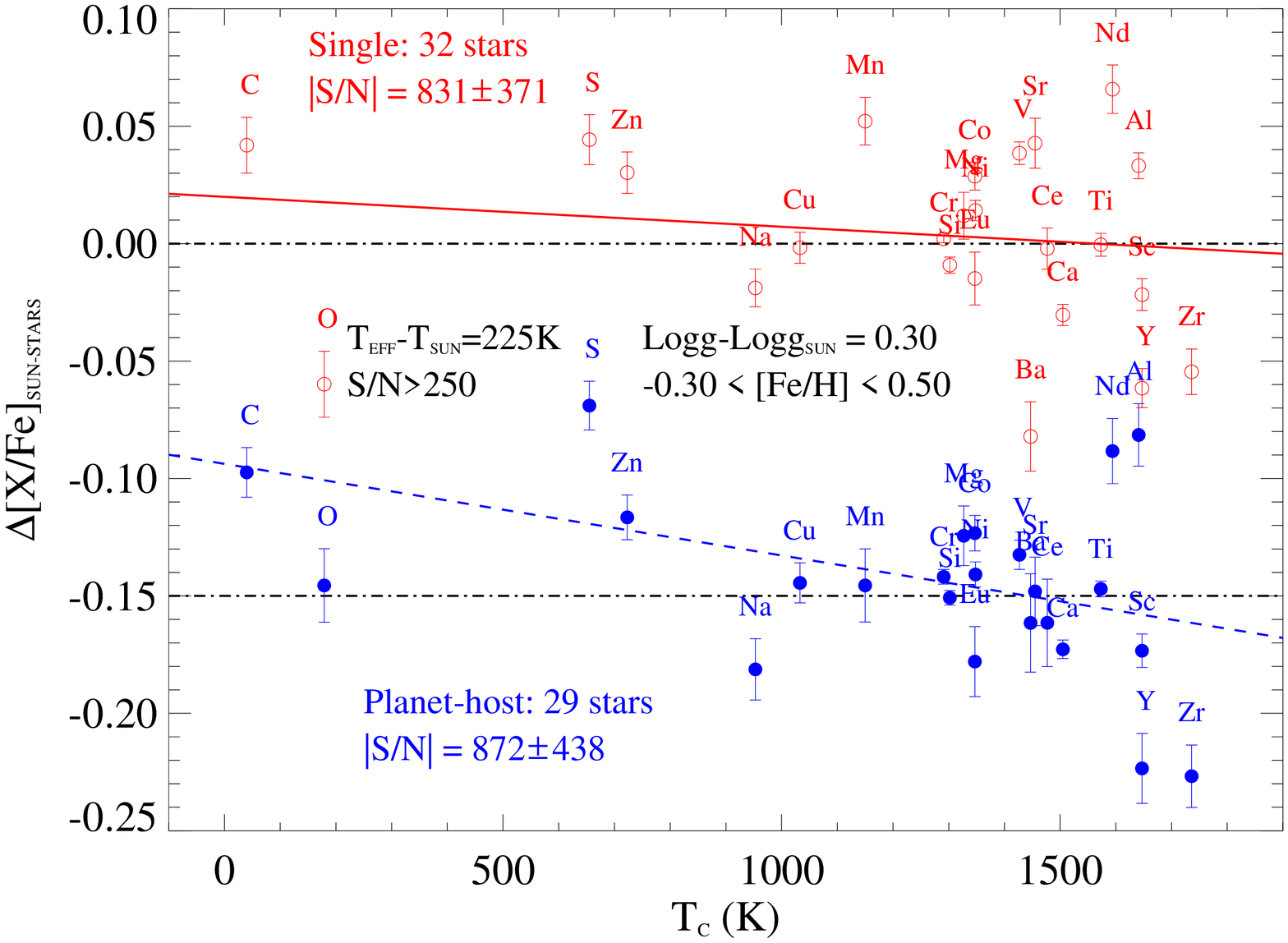}
\includegraphics[width=6.7cm,angle=90]{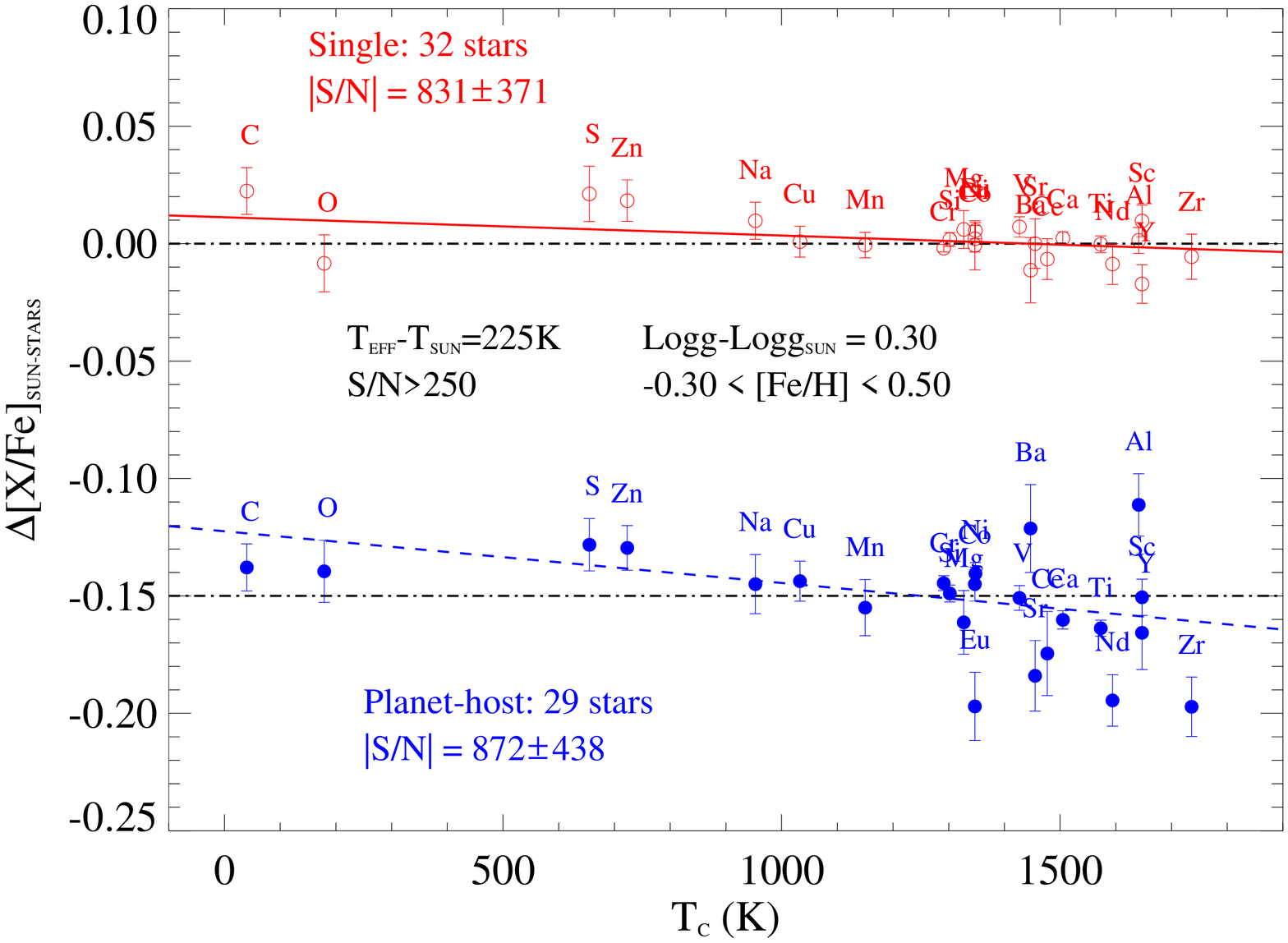}
\caption{
Mean abundance differences, $\Delta {\rm [X/Fe]_{SUN-STARS}}$, 
between the Sun, 29 planet hosts (blue filled circles), and 32 
``single'' stars (red open circles) of the sample of all ``hot'' 
analogs. 
Error bars are the standard deviation from the mean divided by 
the square root of the number of stars. 
The points of stars hosting planets have been artificially 
shifted by --0.15~dex for the sake of clarity. Linear fits to
the data points weighted with the error bars are also displayed 
for planet hosts (dashed line) and ``single'' stars (red solid line). 
The right panel shows the results after removing the galactic 
chemical evolution effects using the linear fits to the 32 
``single'' stars.
}
\label{fha}
\end{figure*}

\section{Sample\label{secsample}}

With the aim of performing a detailed chemical analysis, we 
selected late F- and early G-type stars with spectra at 
S/N~$>250$ from these three spectrographs. 
We end up with a sample of
29 planet-host stars and 32 ``single'' stars (see Fig.~\ref{fpar}). 
All UVES and UES stars are planet hosts. 
In Table~\ref{tblsam} we provide the ranges of stellar parameters 
and metallicities of the 61 ``hot'' analogs of the sample. 
In Fig.~\ref{fpar}, we display the histograms of the effective
temperatures, surface gravities, and metallicities of this sample.
The stars hosting planets are on average more
metal-rich than the stars without planets. 
Thus, we define a subsample of metal-rich late F- and early 
G-type stars with $0.04 < [{\rm Fe}/{\rm H}] < 0.19$, containing 13
planet hosts and 13 ``single'' stars (see Table~\ref{tblsam}). 
On the other hand, due to the recently discovered new planets
\citep[see e.g.][]{may11} in the sample of solar analogs
presented in \citet{gon10}, there are now 33 stars with
planets and 62 stars without planets.

\section{Stellar parameters}

The stellar parameters and metallicities of the HARPS stars
were extracted from the work by \citet{sou08}, which are  
based on the equivalent widths (EWs) of 263 \ion{Fe}{i} and 
36~\ion{Fe}{ii} lines, measured with the code ARES\footnote{The ARES
code can be downloaded at http://www.astro.up.pt/}~\citep{sou07}
and which evaluates the excitation and ionization equilibria. 
For the UVES and UES stars we decided to recomputed the stellar
parameters (see Table~\ref{tblpar}) using these higher quality data. 
Chemical abundances of Fe lines were derived using the 2002 
version of the LTE code MOOG~\citep{sne73}, and a grid of Kurucz 
ATLAS9 plane-parallel model atmospheres~\citep{kur93}.

\section{Chemical abundances\label{secabun}}
 
We performed a fully differential analysis that is, at least, 
internally consistent. Thus, we used the 
HARPS solar spectrum\footnote{The HARPS solar spectrum can be 
downloaded at \\
http://www.eso.org/sci/facilities/lasilla/instruments/harps/
inst/monitoring/sun.html} of the \emph{Ganymede}, 
a \emph{Jupiter}'s satellite, as solar reference \citep[see][]{gon10}. 
We note here that the UVES and UES stars
were also analyzed using the HARPS \emph{Ganymede} spectrum as solar
reference owing to the lack of an appropriate solar spectrum observed
with these two instruments.

We computed the EW of each spectral line using the code
ARES~\citep{sou07} for most of the elements. However, 
as in \citet{gon10}, for O, S, and Eu, we calculated
the EW by integrating the line flux and for Zr, 
by performing a Gaussian fit taking into account possible 
blends. In both cases, we use the task {\scshape splot} within the 
IRAF package. The EWs of Sr, Ba, and Zn lines were checked 
within IRAF, because, as for the elements O, S, Eu, and Zr, 
for some stars, ARES did not find a good fit due to numerical 
problems and/or bad continuum location. 
For further details see Sect.~5 in \citet{gon10}. 
 
The oxygen abundance was derived from the forbidden [\ion{O}{i}]
line. This line is serverely blended with a \ion{Ni}{i} line
\citep[see e.g.][]{all01}. We adopted the atomic parameters for
the Ni line from \citet{all01} and for [\ion{O}{i}] line from 
\citet{lam78}. For more details see \citet{ecu06}. We derived
the EW of Ni line using the mean Ni abundance of the star and
subtracted this EW from the overall EW of the [\ion{O}{i}] 
feature. The remaining EW value give us the O abundance of the star.
In the solar case, the atlas solar spectrum \citep{kur84},
the whole feature has an EW of $\sim 5.4$~m{\AA},
which provides an abundance\footnote{A(X)$=\log [N({\rm
X})/N({\rm H})]+12$} of A(O)~$=8.74$~dex
\citep[see][for more details]{gon10}.

We again used MOOG~\citep{sne73} and ATLAS models~\citep{kur93}
to derive the abundance of each line and thus determined the 
mean abundance of each element relative to its solar abundance on
a line-by-line basis. 
However, to avoid problems with incorrect EW measurements of some 
spectral lines, we rejected all the lines with a different abundance 
from the mean abundance by more than a factor of 1.5 
times the rms. We also checked that we get the same results when 
using a factor of 2 times the rms, but we decided to stay in a
more restrictive position. 
For 34 F- and G-type stars without planets in our sample, 
the line-by-line scatter in the differential abundances goes on 
average from 
$\sigma \sim 0.014-0.025$ for elements such as Si, Ni, Cr, Ca, and Ti, 
to $\sigma \sim 0.035$ for C and Sc, to $\sigma \sim 0.050$ for Mg 
and Na, and to $\sigma \sim 0.063$ for S. These numbers are comparable
to those found in solar analogs \citep[see][]{gon10}.

\section{Discussion\label{secdisc}}

In this section we inspect the abundance ratios of 
different elements as a function of the metallicity, [Fe/H], 
as well as the relation, for different samples of ``hot'' analogs, 
between the abundance difference, 
$\Delta {\rm [X/Fe]_{SUN-STARS}}$, and 
the 50\% equilibrium condensation temperature, 
$T_C$~\citep[see][]{lod03}. We also explore the abundance 
pattern of some particular stars containing super-Earth-like 
planets in the sample of both ``hot'' analogs and solar analogs. 

\begin{figure*}[!ht]
\centering
\includegraphics[width=6.7cm,angle=90]{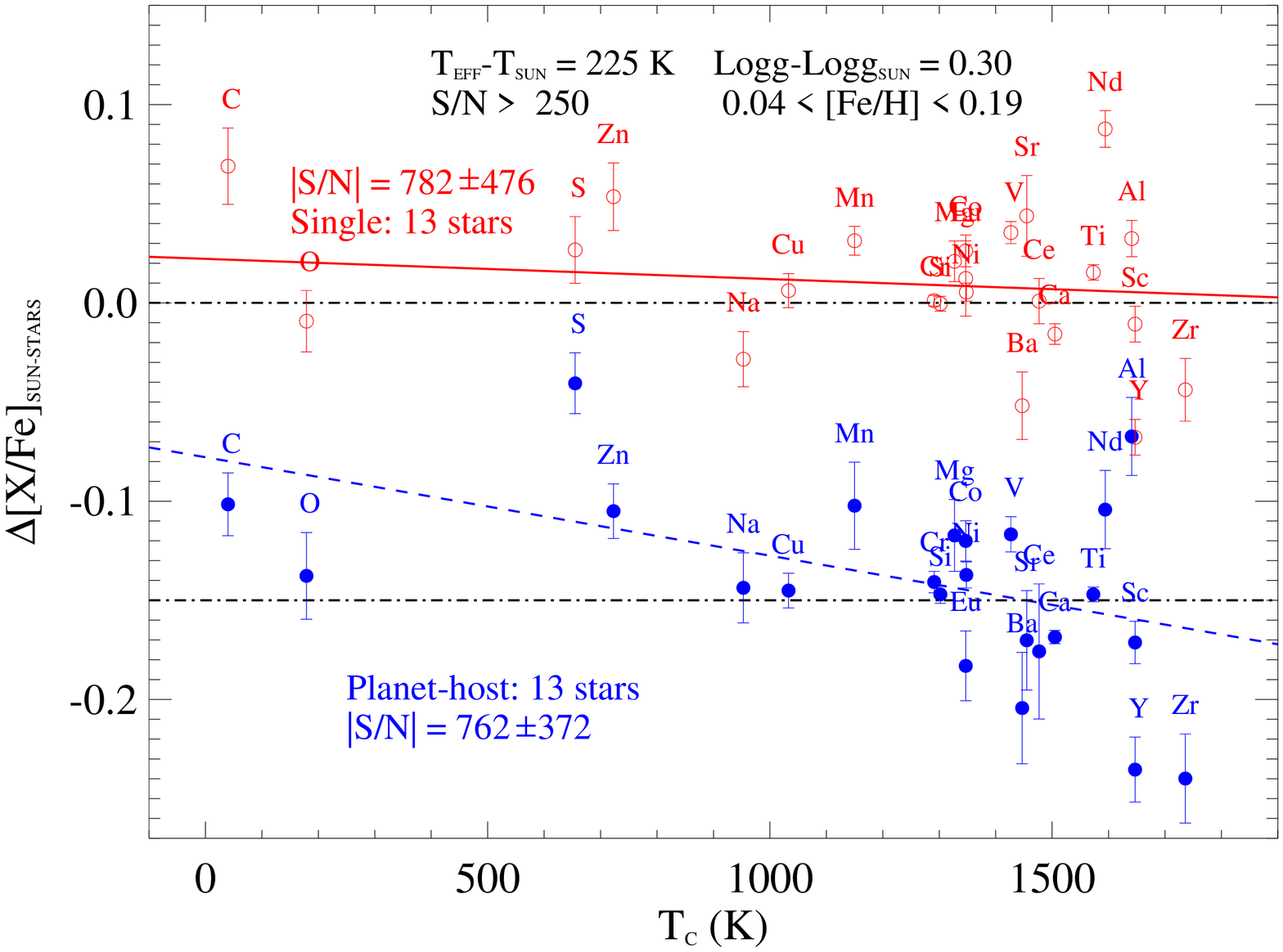}
\includegraphics[width=6.7cm,angle=90]{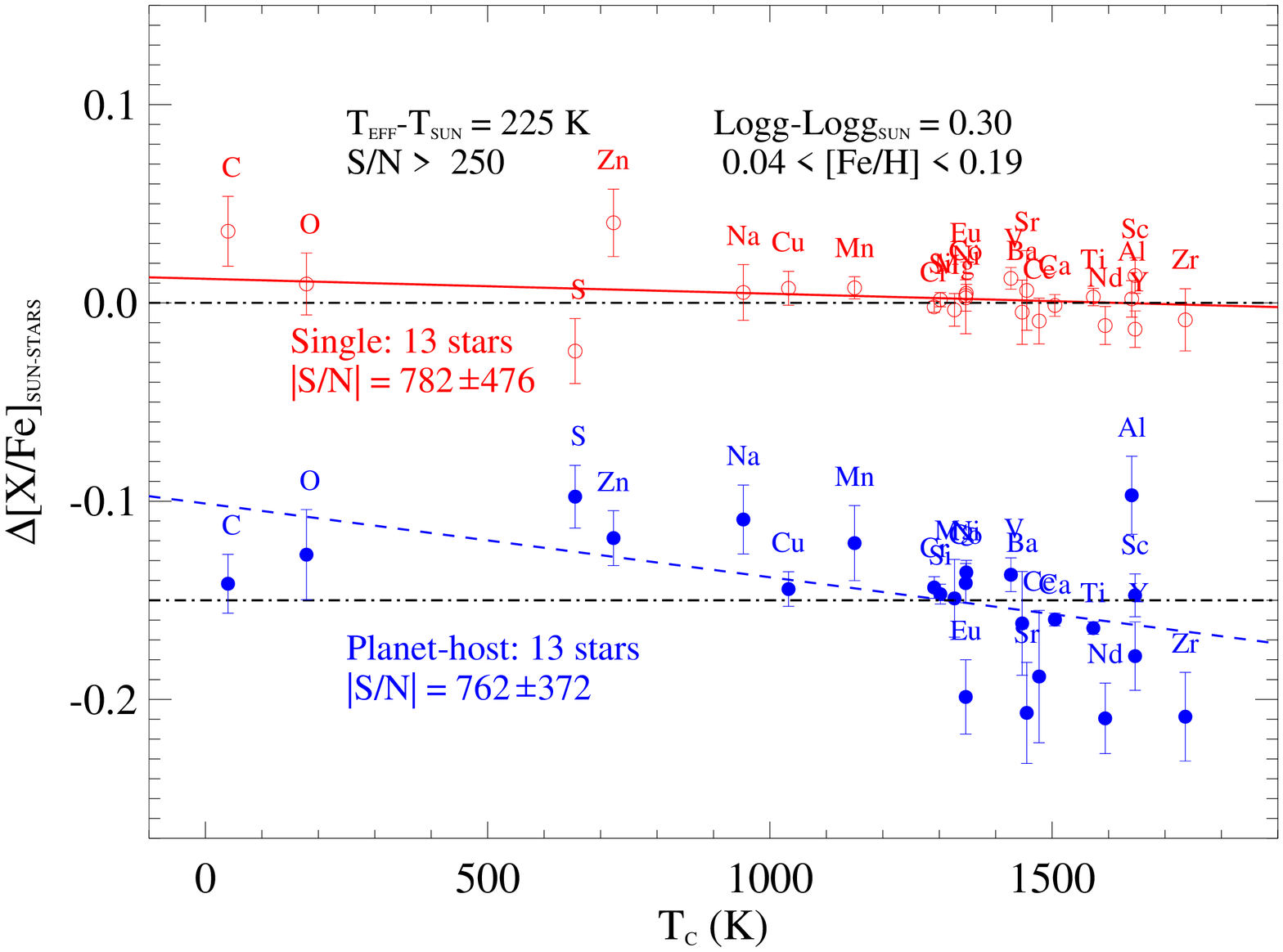}
\caption{ 
Same as in Fig.~\ref{fha} for 13 planet hosts and 13 ``single'' 
stars of the sample of metal-rich ``hot'' analogs.  
}
\label{fmrha}
\end{figure*}

\subsection{Galactic abundance trends\label{secgal}}

In Figs.~\ref{fgala}, and~\ref{fgalc} we 
depict the very accurate 
abundance ratios [X/Fe] of many elements in this sample of late 
F- and early G-type stars with and without planets, derived from 
the high-quality HARPS, UVES, and UES spectra. 
These Galactic trends are compatible with those of Galactic thin disk 
stars~\citep[see e.g.][]{ben05,red06,gil06,tak07,nev09,del10,adi12b} 
and, in particular, with those of solar analogs~\citep{ram09,gon10}. 
We may note the small 
differences in the fits to the abundances of the ``single'' ``hot'' 
analogs and ``single'' solar analogs. In particular, the element
O reveals slight differences at the lowest metallicities (i.e. at
[Fe/H]~$\sim -0.3$ in this sample) whereas Mn, Co, Sr, Ba, Ce and Nd 
present these small differences at the highest metallicities (i.e. at  
[Fe/H]~$\sim +0.3$ in this sample). There is also a tiny shift of 
$\sim 0.03-0.05$ in the fits of Al and V. However, in general for the
rest of the elements like S, Si, Ca, Ti, Cr, and Ni, we almost find
perfect agreement. In Tables~A.1--A.8,  
we provide the element abundance ratios [X/Fe] which are all 
available online\footnote{The online Tables~A.1--A.8
are available in electronic form at the CDS via anonymous ftp to 
cdsarc.u-strasbg.fr (130.79.128.5) or via 
http://cdsarc.u-strasbg.fr/viz-bin/qcat?J/A+A/...}. 



\subsection{Mean abundance ratios [X/Fe] versus $T_C$: previous
results\label{secpp}}

\citet{mel09} suggest that the formation of rocky planets in the 
solar system affects the composition of solar photosphere by 
lowering the amount of refractory elements with respect to the 
volatile elements. They focus on Earth-like planets and 
do not comment on how much the formation of other slightly more 
massive planets like super-Earth- and Neptune-like planets would 
affect their interpretation of the mean abundance pattern of
solar twins with respect to the Sun.

According to the element abundances of the Earth~\citep{kar93},
the bulk composition of a terrestrial 
planet should be dominated by O, Fe, Mg, and Si, with small 
amounts of Ca and Al and with very little C. These terrestrial
planets would be composed by Mg silicates and metallic Fe, 
with other species present in relatively minor amounts.
Water and other hydrous phases may or may not be 
present~\citep{bon10i}.
The composition of Earth-like planets depends on the chemical
composition of the host star~\citep{bon10i,bon12,del10}, and a 
wide variety of terrestrial planets compositions is predicted.
Also giant planet migration plays an important role in the 
final Earth-like compositions~\citep{bon12ii}.
A variation of $\pm 25$~\% from the Earth composition
should be consider to include Venus and Mars as terrestrial 
planets~\citep{bon12}.

Depending of the final compositions, these Earth-like planets
could have substantial amounts of water and develop plate 
tectonics~\citep{bon12ii}, which have been proposed as 
necessary conditions for life.
Simulations of super-Earth-like planets reveal that these slightly
more massive planets could also have plate tectonics~\citep{val07t}.
These authors consider unlikely, although possible, that 
super-Earth-like planets could have a lot of gas 
(Valencia et al. 2007b), as they state that at higher masses 
than 10~\Mea, there is a possibility that H and He 
are retained, although these could give rise to massive ocean 
planets with high H$_2$O content and a relatively small layer of 
gas above it~\citep{val07b}. 
We should note, however, that these simulations of super-Earth-like 
planets assumed a given planetary composition similar to Earth-like 
planets, but it seems reasonable to assume that super-Earth-like 
planets, since formed in the same protoplanetary disk, could 
become a larger counterpart of Earth-like planets and therefore 
have a similar composition.

\citet{ram09} derived the slopes of trends described by the 
abundance ratios of a sample of solar analogs versus the 
condensation temperature, $T_C$. 
Although they only took those elements with $T_C > 900$~K into 
account, they argue that, following the 
reasoning in \citet{mel09}, a negative slope is an indication 
that a star should not have terrestrial planets even if this 
star already contains giant planets. 
We note in this paper, as in \citet{gon10}, that we are evaluating 
the abundance differences $\Delta {\rm [X/Fe]_{SUN-STARS}}$, and 
thus we follow the same reasoning as in \citet{mel09}.
These authors also analyze ten solar analogs, among which four have
close-in giant planets. Their abundance pattern differs from 
the Sun's but resembles that of their solar twins. Although the result
was statistically not significant, they suggest that stars with and 
without close-in giant planets should reveal different abundance 
patterns. 

\citet{gon10} studied the abundance trends of solar analogs 
with and without planets and find that most of these 
stars have negative slopes, hence indicating that these stars 
should not contain terrestrial planets.
Two stars in their sample with super-Earth-like planets, 
HD~1461 and HD~160691, provide abundance trends 
[X/Fe]$_{\rm SUN-STARS}$ versus 
condensation temperature, $T_C$, with the opposite slope to what is 
expected if these stars contain terrestrial planets. 
Therefore, \citet{gon10} suggest that these abundance patterns 
are not related to the presence of terrestrial planets. 
They also note that 
the abundance patterns of solar analogs are strongly affected by 
chemical evolution effects, especially for elements with steep 
Galactic trends like Mn (with $T_C \sim 1160$~K). 

\citet{ram10} combined chemical abundance data of solar analogs from 
previous studies. They produce an average abundance pattern 
consistent with their expectations \citep{ram09} but extracted 
from more noisy abundance patterns. They also argue that when
deriving the slopes for elements with $T_C > 900$~K in the two
planet hosts HD~1461 and HD~160691, they obtained the sign consistent
with the results by \citet{ram09}. \citet{gon11iau} demonstrate
instead that using the elements $T_C > 1200$~K provides the opposite 
sign as in the case of $T_C > 900$~K. Unfortunately, there are more
refractory elements, most of them with $T_C >1200$~K, than volatile 
elements. Thus a more reliable trend should consider the whole range 
of $T_C$ values for the available elements, i.e. $0 < T_C < 1800$~K 
\citep{gon11iau}. 
In addition, subtracting the Galactic chemical evolution effects
before fitting the abundance ratio [X/Fe] almost completely removes 
any trend not from these two planet-host
stars but also from the average values of the subsample of metal-rich
solar analogs \citep{gon11sea}. 

\citet{gui10} and~\citet{gui11} investigated the abundances of 
refractory and volatile elements in a smaller sample of stars 
with and without giant planets. 
They claim to confirm the results by \citet{ram09} but in a 
broader effective temperature, $T_{\rm eff}$, range. 
They also state that more metal-rich stars 
display steeper trends than more metal-poor counterparts. 
\citet{sch11a} analyzed a small sample of ten stars known to host 
giant planets using very high-S/N and very high-resolution spectra 
but in the wide $T_{\rm eff}$ range 5600--6200~K and in the 
metallicity, [Fe/H], range 0-0.4~dex. They chose instead to fit the
abundances [X/H] versus $T_C$ and found positive slopes in four stars
with close-in giant planets and flat or negative slopes in stars with
planets at longer orbital periods, thus consistent with recent
suggestions of the terrestrial planet formation signature. They 
contend, however, that these trends probably result from Galactic 
chemical evolution effects.

More recently, \citet{sch11b} and \citet{ram11} have presented a 
detailed chemical analysis of the two solar twins of the 16 Cygni
binary system. One of them, 16 Cygni B contains a Jupiter-mass planet
in an eccentric orbit, but no planet has been detected around
16 Cygni A. According to \citet{sch11b}, these two stars appear to be
chemically identical, with a mean difference of 0.003$\pm$0.015~dex. 
On the other hand, \citet{ram11} find that star A is more metal-rich
than star B by 0.041$\pm$0.007~dex, and suggested that this may be
related to the formation of giant planet in star B. 
Both results may be consistent at the 3-$\sigma$ level and probably 
indicate that internal systematics are greater than typically assumed.
\citet{tak05} claims there is no metallicity difference between 16
Cygni A and B. \citet{lag01} were the first in performing a 
differential analysis and they also suggest that star A is more
metal-rich than star B by 0.025$\pm$0.009~dex, but only based on the
analysis of iron lines.

\begin{figure*}[!ht]
\centering
\includegraphics[width=6.7cm,angle=90]{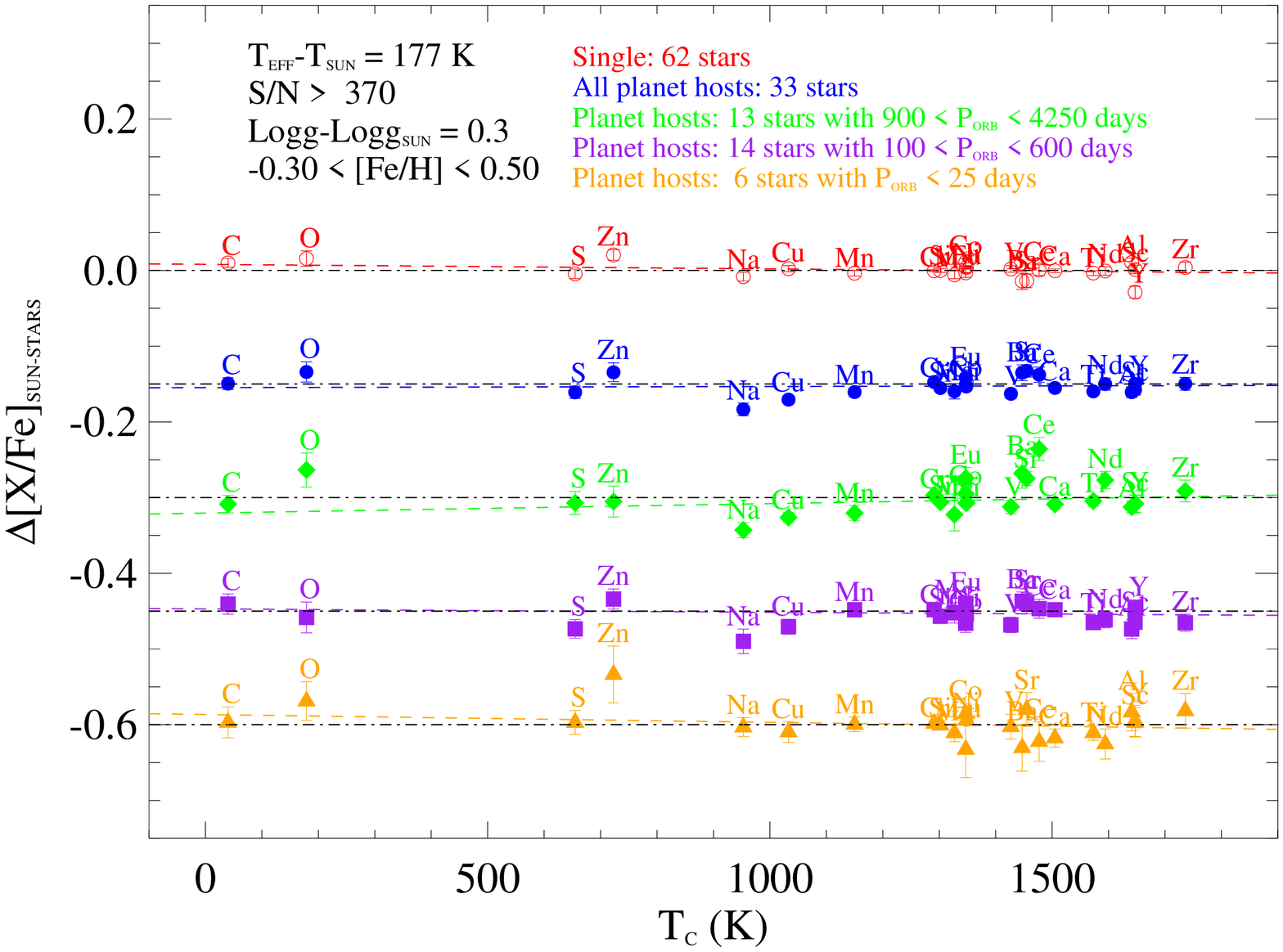}
\includegraphics[width=6.7cm,angle=90]{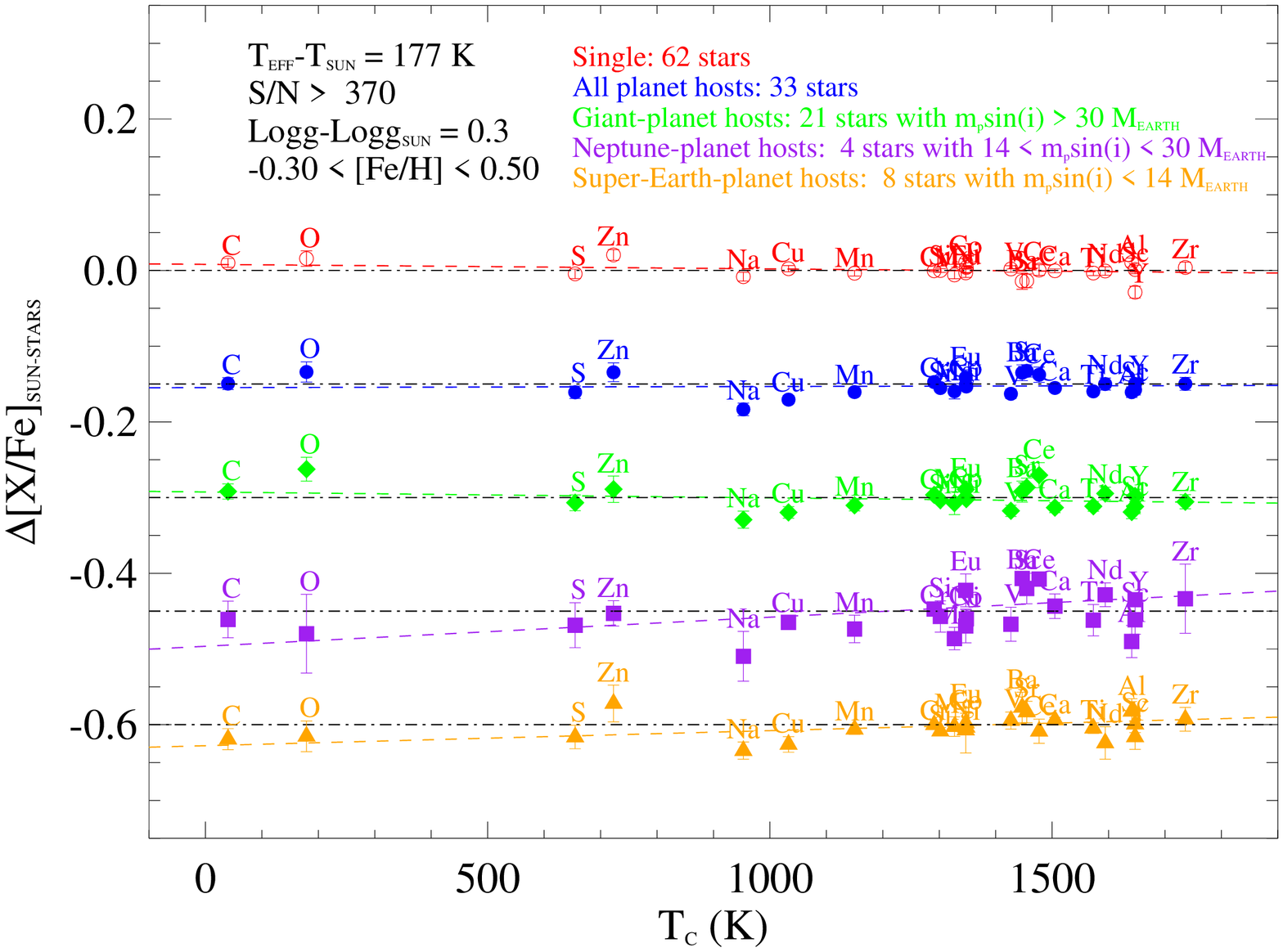}
\caption{{\it Left panel:} Same as right panel 
in Fig.~\ref{fha} but for solar analogs
containing 62 ``single'' stars (red open circles) and 33 stars 
hosting planets (blue filled circles), with the most massive planet 
in an orbital period $P_{\rm orb} < 25$ days (6 stars, yellow 
filled triangles), $100 < P_{\rm orb} < 600$ days 
(14 stars, violet filled squares), $900 < P_{\rm orb} < 4250$ days 
(13 stars, green filled triangles). 
The mean abundance differences, $\Delta {\rm [X/H]_{SUN-STARS}}$, 
and linear fits have been artificially shifted by $-0.15$~dex, for 
the sake of clarity. Horizontal dashed-dot lines show the zero point 
levels for each set of points. {\it Right panel:} Same as left panel
but for planet-host stars with the minimum mass of the least 
massive planet in the ranges: $m_p \sin i < 14$~\Mea (8 stars, yellow 
filled triangles), $14 < m_p \sin i < 30$~\Mea 
(4 stars, violet filled squares), $m_p \sin i > 30$~\Mea 
(21 stars, green filled triangles).}
\label{fsa}
\end{figure*}

\subsection{All ``hot'' analogs\label{secall}}

The whole sample of stars presented in this work is composed of 
61 late F- and early G-type stars, 29 planet hosts, and 32 ``single''
stars. 
In Fig.~\ref{fha} we display the mean abundance difference, 
$\Delta {\rm [X/Fe]_{SUN-STARS}}$, versus 
the condensation temperature, $T_C$, using the HARPS  
spectrum\footnote{The HARPS solar spectra can be downloaded at 
http://www.eso.org/sci/facilities/lasilla/instruments
/harps/inst/monitoring/sun.html} of \emph{Ganymede} 
as solar reference \citep[see][for further details]{gon10}.
The left panel of Fig.~\ref{fha} evokes high scatter among 
the points for different elements for stars both with and without
planets, and therefore the fits are very tentative and do not 
provide any additional information. However, that in 
this relatively narrow metallicity range the Galactic trends
shown in Figs.~\ref{fgala}, and~\ref{fgalc} give 
different 
slopes, strongly affects the mean element abundance ratios.
For this reason, we also computed the values 
$\Delta {\rm [X/Fe]_{SUN-STARS}}$ but after subtracting these 
Galactic evolution effects. To do that, we fit a linear function to
the Galactic trends of the 32 ``hot'' analogs without planets 
(see linear fits in Figs.~\ref{fgala}, and~\ref{fgalc}) 
and subtracted the value given by this fit at the metallicity of each 
star to the abundance ratio of the star. 

The result is depicted in the right-hand panel of
Fig.~\ref{fha}. For ``single'' stars, the mean 
abundance ratios of pratically all elements roughly become zero, 
with a tiny scatter ($\sigma = 0.0097$),
as well as the slope of the fit to all mean abundances is 
almost zero, with a value of 
$(-0.09\pm0.04)\times 10^{-4}$~dex~K$^{-1}$. 
For ``hot'' analogs hosting planets, the remaining scatter is larger, 
so the slope of the fit remains negative, 
$(-0.24\pm0.05)\times 10^{-4}$~dex~K$^{-1}$ (see
Table~\ref{tblslope}). 
This is probably because we are using the linear fits 
to the Galatic trends of stars without planets,
although the abundance ratios of planet hosts seem to 
follow the Galactic trends defined by these linear fits. 
However, if we choose planet hosts to fit their galactic chemical 
trends the result give a similar scatter ($\sigma = 0.0121$) as the
that that abundance ratios of ``single'' stars in Fig.~\ref{fha}, 
and the fit is close to zero ($-0.013\pm0.046$).
Hereafter we use abundance ratios of ``single'' stars to fit the
Galactic trends because they cover the whole range of metallicity, 
whereas planet hosts mostly cover metallicities [Fe/H]$ > 0$, 
thus ``single'' stars are better tracers of Galactic chemical 
evolution effects.

This may be also due to intrinsic differences between the abundance 
pattern of stars with and without planets, a possibility that we 
explore in Section~\ref{secstp}. 
This could be also related to the different metallicity 
distributions of stars with and without planets, where the
former is more metal-rich, but as we see in Section~\ref{secmr}, a
subsample of metal-rich ``hot'' analogs without planets still 
behaves in a different way than a similar sample of stars 
with planets.    

\begin{figure*}[!ht]
\centering
\includegraphics[width=12.5cm,angle=90]{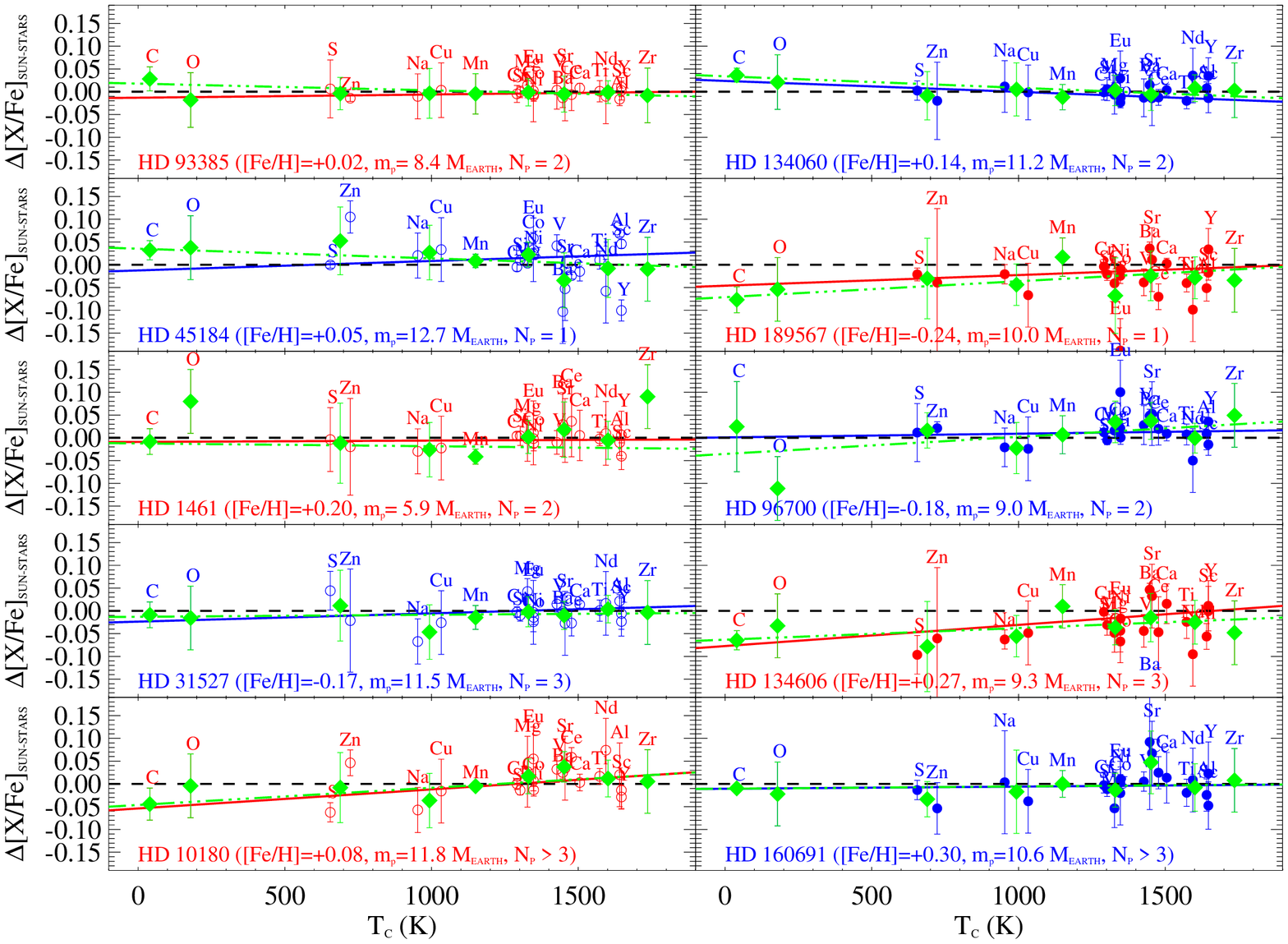}
\caption{
Abundance differences, $\Delta {\rm [X/Fe]_{SUN-STARS}}$, 
between the Sun and 10 stars hosting super-Earth-like planets 
(circles), 2 corresponding to the sample of ``hot'' analogs 
(top panels) and 8 belong to the sample of solar analogs 
analyzed in \citet{gon10}. 
Error bars are the uncertainties of the element abundance 
measurements, corresponding to the line-by-line scatter. 
Diamonds show the average abundances in bins of $\Delta T_C = 150$~K. 
Error bars are the standard deviation from the mean abundance of
the elements in each $T_C$ bin.
Linear fits to the data points (solid line) and to the mean data 
points (dashed-dotted line) weighted with the error bars are 
also displayed.
}
\label{fsep}
\end{figure*}

\begin{table}
\caption{Slopes of the linear fit to the mean values 
$\Delta {\rm [X/Fe]_{SUN-STARS}}$ versus $T_C$}
\centering          
\begin{tabular}{lrrr}
\noalign{\smallskip}
\noalign{\smallskip}
\noalign{\smallskip}
\hline\hline       
\noalign{\smallskip}
Sample$^{\rm a}$ & Non-GCE$^{\rm b}$ & 
GCE$^{\rm b}$ & $N_{\rm stars}$$^{\rm c}$ \\  
\noalign{\smallskip}
\hline\hline       
\noalign{\smallskip}
sSA    & $-0.378\pm0.033$ & $-0.060\pm0.029$ & 62 \\
pSA    & $-0.302\pm0.047$ & $+0.014\pm0.040$ & 33 \\

smrSA  & $-0.235\pm0.055$ & $-0.062\pm0.053$ & 16 \\
pmrSA  & $-0.194\pm0.048$ & $-0.027\pm0.047$ & 16 \\

sHA    & $-0.160\pm0.046$ & $-0.086\pm0.041$ & 32 \\
pHA    & $-0.404\pm0.048$ & $-0.242\pm0.046$ & 29 \\

smrHA  & $-0.111\pm0.062$ & $-0.106\pm0.061$ & 13 \\
pmrHA  & $-0.506\pm0.066$ & $-0.379\pm0.063$ & 13 \\

smrHAh & $-0.330\pm0.069$ & $-0.258\pm0.070$ & 10 \\
pmrHAh & $-0.419\pm0.068$ & $-0.289\pm0.067$ & 10 \\
 
pSAlp & $-0.099\pm0.067$ & $+0.125\pm0.059$ & 13 \\
pSAmp & $-0.370\pm0.075$ & $-0.044\pm0.062$ & 14 \\
pSAsp & $-0.404\pm0.114$ & $-0.103\pm0.092$ & 6 \\

pSAlm & $-0.348\pm0.059$ & $-0.077\pm0.050$ & 21 \\
pSAmm & $+0.236\pm0.114$ & $+0.383\pm0.097$ & 4 \\
pSAsm & $-0.223\pm0.084$ & $+0.199\pm0.065$ & 8 \\
\noalign{\smallskip}
\hline                  
\noalign{\smallskip}
\noalign{\smallskip}
\end{tabular}

\tablefoot{Slopes of the linear fit of the mean abundance ratios,
$\Delta {\rm [X/Fe]_{SUN-STARS}}$, as a function of the condensation
temperature, $T_C$, using the whole $T_C$ interval, for different 
stellar samples.
\tablefoottext{a}{Stellar samples: ``single'' solar analogs, ``sSA'',
planet-host solar analogs, ``pSA'', ``single'' metal-rich solar
analogs,  ``smrSA'', planet-host metal-rich solar analogs, ``pmrSA''. 
``HA'' refers to ``hot'' analogs and ``HAh'' refers to very high S/N 
data (S/N~$>550$).
Planet-host solar analogs with the most massive planet in an 
orbital period $P_{\rm orb} < 25$ days, ``pSAsp'', 
$100 < P_{\rm orb} < 600$ days , ``pSAmp'', 
$900 < P_{\rm orb} < 4250$ days, ``pSAlp''. 
Planet-host solar analogs with the minimum mass, $m_p\sin i$,
of the least massive planet in the ranges: $m_p < 14$~\Mea , ``pSAsm'',
$14 < m_p < 30$~\Mea, ``pSAmm'', $m_p > 30$~\Mea, ``pSAlm''.} 
\tablefoottext{b}{``Non-GCE'' and ``GCE'' refers to the mean 
abundances before and after being corrected for galactic chemical 
evolution effects.}
\tablefoottext{c}{Total number of stars in the sample.}
}
\label{tblslope}      
\end{table}

\subsection{Metal-rich ``hot'' analogs\label{secmr}}

We selected a subsample of 26 metal-rich ``hot'' analogs with 
the aim of investigating the abundance pattern of similar samples 
of stars with and without planets with very similar metallicity
distribution. Thus, we selected 13 planet-host and 13 ``single'' 
late F- and early G-type stars. 
In Fig.~\ref{fmrha} we display the mean values 
$\Delta {\rm [X/Fe]_{SUN-STARS}}$ for these two subsamples. 
In the left-hand panel of Fig.~\ref{fmrha}, planet hosts again yield 
a steeper abundance trend versus condensation temperature than
``single'' hot analogs. 
However, in the right-hand panel of Fig.~\ref{fmrha}, after removing 
the chemical evolution effects, the abundance pattern of ``single'' 
stars does not exhibit any trend whereas for planet hosts 
the mean abundances still hold a negative slope. 

\citet{mel09}, \citet{gui10}, and \citet{gui11} found that 
the planet-host stars show 
steeper trends than ``single'' stars at high metallicity 
although this effect is probably related to chemical evolution
effects. \citet{gon11sea} depicted in their Fig.~3 the mean
abundance trends of a subsample of metal-rich solar analogs with 
and without planets together with steep fits to the data, but with
similar slopes in both cases, 
$(-0.23\pm0.06)\times 10^{-4}$~dex~K$^{-1}$ for ``single'' stars 
and $(-0.16\pm0.05)\times 10^{-4}$~dex~K$^{-1}$ for planet-host 
stars \citep[see also][]{gon10}. 
They display in their Fig.~4 however that after removing the galactic 
chemical evolution effects, these negative trends almost disappear, 
and clearly the slopes of stars with and without planets are
consistent within the 1-$\sigma$ uncertainties, 
$(-0.07\pm0.06)\times 10^{-4}$~dex~K$^{-1}$ for ``single'' stars 
and $(-0.02\pm0.05)\times 10^{-4}$~dex~K$^{-1}$ for planet-host 
stars (see Table~\ref{tblslope}).
In addition, the mean abundance pattern of solar analogs after
correcting for galactic chemical evolution also exhibit flat trends 
for the 62 ``single'' stars and 33 planet hosts \citep{gon13aas}. 
On the other hand, if one chooses a subsample of very high-quality
spectra of metal-rich ``hot'' analogs with S/N~$> 550$, then although
the trends still remain very steep with negative slopes, in this 
case, for ten stars with and ten without planets \citep{gon13an}, 
these slopes are both very similar and consistent within the 
error bars (see Table~\ref{tblslope}).

\begin{figure}[!ht]
\centering
\includegraphics[width=6.7cm,angle=90]{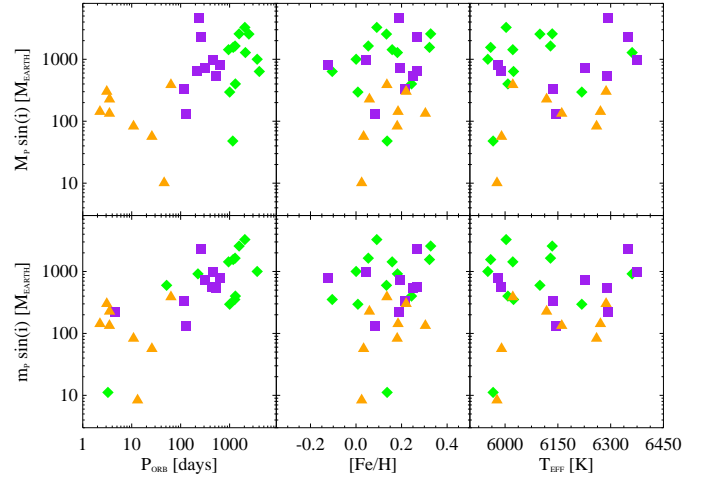}
\caption{
{\it Top panels}: Minimum mass of the most massive planet of 
each of ``hot'' analogs hosting planets versus the orbital 
period of the planet (left panel) and metallicity 
[Fe/H] (middle panel), and temperature (right panel), 
with the most massive planet in an orbital period $P_{\rm orb} < 65$ 
days (diamonds), $115 < P_{\rm orb} < 630$ days (squares), 
$950 < P_{\rm orb} < 4050$ days (triangles).
{\it Bottom panels}: the same as top panels but showing the minimum 
mass of the least massive planet. The symbols as in top panel.
}
\label{fmoha}
\end{figure}

\subsection{Revisiting solar analogs\label{secsa}}

The number of planet hosts in the sample of solar analogs in
\citet{gon10} have changed due to the discovery of planets of
different masses, in particular, low-mass planets with masses below
30~\Mea \citep[see e.g.][]{may11}. We have reanalyzed the data 
presented in \citet{gon10} and been able to correct the
abundance differences, $\Delta {\rm [X/Fe]_{SUN-STARS}}$, for the
chemical evolution effects using the linear fits to the Galactic 
abundance ratios, [X/Fe], of solar analogs without known planets. 
 
In Fig.~\ref{fsa} we display
the mean abundance differences for updated lists of stars with and
without detected planets. Stars hosting planets are also separated
for different orbital periods of the most massive planet in the
planetary systems. However, we do not see any clear signature. All
of them show similar mean abundance pattern although with slightly 
different slopes, to those found for the whole samples of stars 
with and without planets. However, one could speculate that the
positive trend seen for the largest orbital periods
($P_{\rm orb} > 900$ days) would allow the formation of terrestrial 
planets in closer orbits. For the most massive planets at the 
orbital periods between 100 and 600~days, we get a flat trend (see
Table~\ref{tblslope}).
The stars with the most massive planets at the shortest periods 
($P_{\rm orb} < 25$ days) display a slightly negative trend, which
would indicate that these stars would not have formed
terrestrial planets or that they have been captured via, for
instance, a possible inward migration of the most massive planets 
in these systems. A similar slope is found for stars without 
detected planets; however, these slope values are only significant at
$\leq 2$~$\sigma$ (see Table~\ref{tblslope}), which probably makes 
the previous statement too tentative.

In the right-hand panel of Fig.~\ref{fsa}, we depict the mean 
abundance differences but separate the planet hosts into 
several ranges of the
minimum mass of the least massive planet. The mean abundance patterns 
are similar, but the slopes of the trends seem to behave in 
different ways depending on the range of masses. 
We again try to speculate on 
these results to try to identify possible explanations. 
The stars containing super-Earth-like planets show an increasing 
trend, which would indicate that these stars are slightly different
from the Sun and that the super-Earth-like planets behave on average 
like the terrestrial planets of the solar planetary system, but with 
higher amount of metals trapped on these super-Earth-like planets. 
On the other hand, the stars 
harboring Neptune-like planets show on average a steep positive 
trend slope that would mean that the Neptune-like planets have, 
as one may expect, a higher content of refractory elements than 
the Sun and that the volatile content caught in Neptune-like planets 
would not be too large.
Finally, the stars hosting Jovian planets exhibit a slightly negative
or almost flat trend, which may indicate that these stars only 
harbor giant planets, although in that case one would expect to see
a steep negative trend. 
The possibility still remains that these stars harbor a small
number of terrestrial planets.
The ideal situation would be to group the stars taking
the planet mass and periods into account
at the same time but the statistics 
would be even worse. 
Thus, although these statements appear to follow the line of 
reasoning in \citet{mel09}, we think we need more Sun-like stars with
already detected planets at different masses and orbital periods and
with high-quality data to confirm this very tentative scenario. 
In the next section we individually inspect the abundance pattern of
the stars hosting super-Earth-like and Neptune-like planets to try 
to elucidate whether there is a connection between the amount of 
rocky material in the low-mass planets and the slopes of the 
abundance trends of each star.

\subsection{Stars hosting super-Earth-like planets\label{secse}}

\citet{gon10} already studied the abundance trends of two stars in
their sample, with negative slopes, which harbor super-Earth-like 
planets, HD~1461 and HD~160691. 
Taking into account the new discovered planets presented e.g. 
in \citet{may11}, our sample of solar analogs analyzed in 
\citet{gon10} and our sample of ``hot'' analogs now contain 8 
and 2 stars hosting super-Earth-like planets, respectively.
It is therefore interesting to look at the abundance pattern of 
these stars harboring, at least, super-Earth-like planets under the 
assumption that these exoplanets contain a significant amount of 
rocky material in the form of refractory elements. 
We note here that the masses of the planet are indeed minimum
masses, so that in some cases the ``real'' planet mass could 
be significantly higher.
However, statistical analysis show that the distribution of $m_p \sin
i$ values is similar to that of ``$m_p$'' values~\citep{jor01}. 
In fact, the average value of $\sin i$ for a random distribution of
angles is of 0.79, so the average factor of overestimation is 
only 1.27.

In Fig.~\ref{fsep} we display the abundance pattern of ten  
stars hosting super-Earth-like planets in the sample of ``hot''
analogs and the sample of solar analogs analyzed in \citet{gon10}. 
The abundance differences of these stars, 
$\Delta {\rm [X/Fe]_{SUN-STARS}}$, are plotted versus the 
condensation temperature, $T_C$. 
The abundances ratios of each star were corrected for the Galactic 
trends at the metallicity of the star. To compute these abundance 
corrections, we used the linear fitting functions of the galactic
abundances of both the sample of 32 ``hot'' and 62 solar analogs 
without detected planets. 
In Fig.~\ref{fsep} we also show the fits to all abundance ratios 
of each star weighted by their uncertainties. 
There are many more refractory elements (with $T_C \gtrsim 1200$~K) 
than volatile elements, so a more reliable fit of these
element abundance ratios should consider all of the $T_C$ values. 
However, we still find that this fitting
procedure does not give the same weight to all $T_C$ values and 
thus we decided to average the abundance ratios in bins of 
$\Delta T_C = 150$~K. 
We estimate the error bars of these new points as the maximum value
between standard deviation from the mean abundance and the mean
abundance uncertainty of the elements at each $T_C$ bin, 
unless there is only one point in this $T_C$ bin, in which case, 
the error bar is associated with the
abundance uncertainty of the given element. 
The error bar is estimated in this way in order to consider 
all possible sources of uncertainty, i.e. not only the 
scatter of the element abundances but also the individual abundance
uncertainties.
These new data points are also depicted in Fig.~\ref{fsep},
together with their corresponding linear fitting functions, which 
were again derived by the fit of these points weighted by 
their error bars. 
The slope of these linear fits is our adopted slope to be evaluated 
in the context of the possible presence of terrestrial planets in
these stars.
In many cases, these new fits of the average points are similar to 
those corresponding to the individual element abundances.  

\begin{figure}[!ht]
\centering
\includegraphics[width=6.7cm,angle=90]{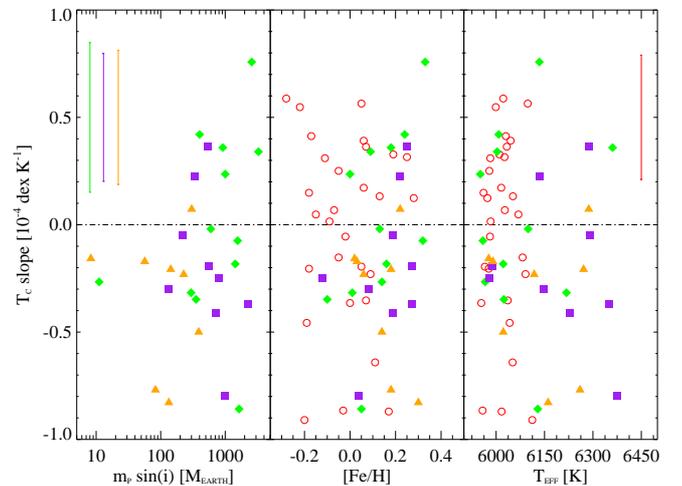}
\caption{
Slopes of the linear fits to 61 ``hot'' analogs
of the mean element abundance ratios,
 $\Delta {\rm [X/Fe]_{SUN-STARS}}$,
versus the condensation temperature, $T_C$, as a function
of the minimum mass of the least massive planet of each 
planet-host star (left panel),
the stellar metallicity (middle panel), and effective temperature
(right panel). 
The mean error bars of the slopes are shown in 
upper-left corner for planet hosts and upper-right corner for
``single'' stars. The symbols as in Fig.~\ref{fmoha}.
}  
\label{fslpha}
\end{figure}

By inspecting Fig.~\ref{fsep}, one realizes that there are four stars
showing positive slopes, three showing negative slopes, and three
giving a flat slope. We note that if all the stars provided a 
flat slope, the interpretation related to the existence 
of terrestrial planets would lose its meaning. 
Either way, the magnitude of the slope
should be related to the amount of rocky material 
present in these planetary systems. Thus, finding one single 
planetary system that does not accomplish the expected slope may be
enough to rule out the statement claiming that the abundance 
pattern of a star conceals a signature of rocky planets. In these 
small subsamples of ten stars hosting small planets, there are three
stars providing negative trends, but having relatively massive
super-Earth-like planets. 
In these three stars, one would expect to find,
at least, similar positive trends. 
However, we still do not know if these stars actually 
harbor really rocky Earth-like planets. 

In Fig.~\ref{fsep} we also indicate the number of confirmed planets 
in these planetary systems. 
There does not seem to be any connection between the number of planets
and the slope of the abundance pattern versus condensation
temperature. There are solar analogs with only two detected planets, 
HD~1461 and HD~96700, whose less massive planet have 
5.9~\Mea and 9.0~\Mea at orbital periods 
$P_{\rm orb}\sim$~13.5~days and~8.1 days, respectively. These two 
solar analogs also have a second super-Earth-like planet with masses 
7.6~\Mea and 12.7~\Meao, at $P_{\rm orb}\sim5.8$~days and 103.5~days, 
respectively, hence a significant amount of rocky material. 
Nevertheless, they both exhibit unexpected, nearly flat trends. 
In addition, the two ``hot'' analogs, 
with $T_{\rm eff}\sim 5975$~K, should have narrower convective 
zones than the Sun what would make it easier to detect any possible 
signature for terrestrial planets. These two stars, HD~93385 and 
HD~134060, harbor two planets each with masses 8.4~\Mea and 10.2~\Mea 
(at $P_{\rm orb}\sim$~13.2 and 46 days) and 11.2~\Mea and 
47.9~\Mea (at $P_{\rm orb}\sim$~3.2 and 1160.9 days). 
These planets should also contain a large amount of rocky material, 
hence refractory elements, but they have slightly lower 
volatile-to-refractory abundance ratios than the Sun, and therefore
should have less rocky material than the solar planetary system.

\begin{figure}[!ht]
\centering
\includegraphics[width=6.7cm,angle=90]{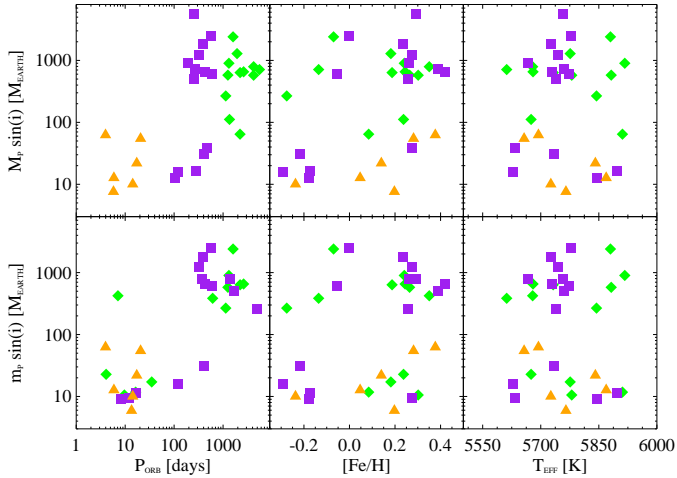}
\caption{ 
Same as Fig.~\ref{fmoha}, but for solar analogs hosting planets, 
with the most massive planet in an orbital period $P_{\rm orb} < 25$ 
days (diamonds), $150 < P_{\rm orb} < 650$ days (squares), 
$1000 < P_{\rm orb} < 4300$ days (triangles).
}  
\label{fmosa}
\end{figure}

It is actually plausible that many of our stars in both samples of
``hot'' and solar analogs host really terrestrial planets. 
Numerical simulations reveal that low-mass (from Neptune-like to
Earth-like) planets are much more common than giant planets, this 
effect is more significant as the metallicity decreases, and probably 
80--90\% of solar-type stars have terrestrial 
planets \citep[see e.g.][]{mor09a,mor09b,mor12}. 
This statement agrees with the growing population of low-mass planets
found in the HARPS sample of exoplanets \citep{udr07,how10,may11}.

\begin{figure}[!ht]
\centering
\includegraphics[width=6.7cm,angle=90]{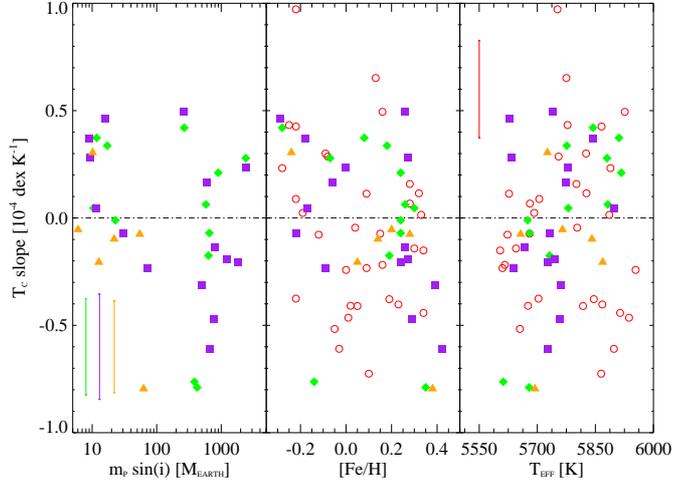}
\caption{
Same as Fig.~\ref{fslpha} but for solar analogs. 
The symbols as in Fig.~\ref{fmosa}. 
}
\label{fslpsa}
\end{figure}

\subsection{Planetary signatures in the abundance trends?\label{secstp}}

We now explore the abundance trends of the whole sample of ``hot'' 
and solar analogs to extract information on the amount of rocky 
material in their planetary systems. In Fig.~\ref{fmoha} we depict 
the important orbital properties, minimum mass of the most massive 
and the least massive planets, and the orbital period as well as 
two stellar properties, metallicity, and effective temperature, 
of all the 29 ``hot'' analogs with planets. 
Figure~\ref{fmoha} exhibits a variety of
planetary systems, many of them with more than one or two planets.
This plot also allows us to see that there are only two stars hosting
super-Earth-like planets (see also Fig.~\ref{fsep}) with 
$T_{\rm eff}\sim 5975$~K and metallicities slightly above solar, 
but also that the most massive planets in these systems are 
another super-Earth-like and one Saturn-like planet in about 6-~and
100-day orbits (see Section~\ref{secse}). In Fig.~\ref{fmoha} we
are also able to distinguish many stars harboring isolated planets
with masses between 50 and 3500~\Mea. The stellar
properties of the stars, $T_{\rm eff}$ and [Fe/H], exhibit  
uniform behavior irrespective for the orbital period of the most 
and the least massive planets in each planetary system. 

Although the planetary systems of the sample of ``hot'' analogs 
exhibit a substantial complexity, we try to extract information 
from the slopes of the average abundances versus the condensation
temperature as seen in Section~\ref{secse}. Thus we derive the 
trend in the average abundances in $T_C$ bins of each ``hot'' 
analog as shown for the stars hosting super-Earth-like planets in 
Fig.~\ref{fsep}.
In Fig.~\ref{fslpha} we depict the slopes of these average trends 
against the minimum mass of the least planet mass, the
stellar metallicity, and the effective temperature. We also display 
the slopes for ``hot'' analogs without known planets. According to
the line of reasoning in \citet{mel09}, positive slopes would mean 
that these stars are deficient in refractory elements with respect 
to the Sun and hence likely to host terrestrial planets. From this 
plot one realizes that only a few stars that already harbor giant
planets would also have terrestrial planets. However, most of the
``hot'' analogs, in particular two stars that already contains 
super-Earth-like planets exhibit negative slopes, so have less
probability of hosting terrestrial planets. We also include stars
without known planets and more than half of them show positive 
slopes and therefore should have terrestrial planets.

In Fig.~\ref{fmosa}, we display the minimum masses of the planets
orbiting 33 solar analogs similarly as Fig.~\ref{fmoha}. 
We notice here that there are eight solar analogs with 
super-Earth-like 
planets with masses between 6~\Mea and 13~\Mea at orbital periods 
between 6 and 16 days, and four Neptune-like planets, with masses 
16~\Mea and 23~\Mea at $P_{\rm orb}$ in the range 4--122~days.  
The distributions of effective temperatures and metallicities among
solar analogs are also approximately homogeneous and do not 
depend on the orbital period of the planets. 
In Fig.~\ref{fslpsa}, we depict the slopes for solar analogs with and
without planets computed as in Fig.~\ref{fslpha}. We also display the
mean error bars for different samples of stars with and without
planets to illustrate how accurate these slopes are. In fact, many of
these values are consistent with flat slopes. 
We see that many stars with low-mass planets exhibit 
positive slopes. 
Most of these planetary systems have Jupiter-like or Saturn-like 
planet at orbital periods longer than 100~days. On the other hand, 
the two solar analogs, showing negative slopes, only harbor 
super-Earth-like planets at orbital periods shorter than 25~days.
Solar-type stars with only giant planets with long orbital periods 
display mostly negative slopes, which may indicate that they do not
host terrestrial planets, although there are some with positive slopes. 
The solar analogs without detected planets are more or less equally
distributed above and below the flat slope.

\begin{figure}[!ht]
\centering
\includegraphics[width=6.7cm,angle=90]{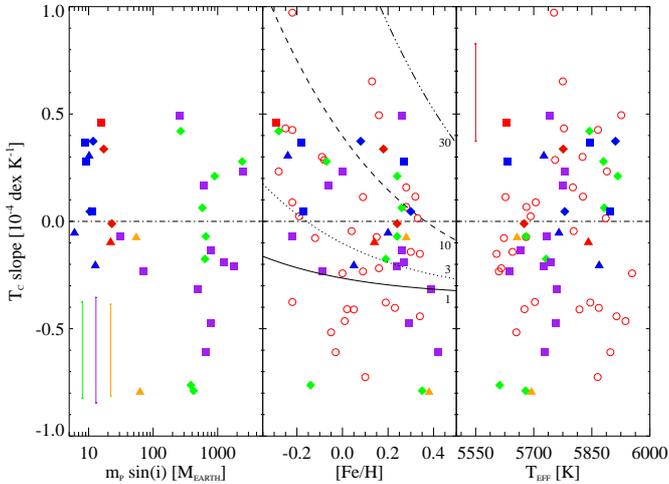}
\caption{ 
Same as Fig.~\ref{fslpsa} but highlighting stars with 
super-Earth-like planets (red filled symbols) and 
stars with Neptune-like planets (blue filled symbols). 
The predicted lines (black lines) track the 
amount of rocky material in the planetary systems as 1, 3, 
10, and 30 \Meao.
}
\label{fslpsat}
\end{figure}

\citet{cha10} demonstrates that is possible to explain the abundances
of solar twins in \citet{mel09} by adding four Earth masses of rocky
material. In the solar system, the terrestrial planets only have 
about 2~\Mea of rock. 
However, the primordial asteroid belt
probably contained a similar amount of rocky protoplanets
\citep{wei77}. This mass of rock was probably ejected from the solar
system once Jupiter formed \citep[see e.g.][]{cha01}, and thus would
be missed from the solar photosphere. 
Therefore, the existence of the present
terrestrial planets and the asteroid belt in the solar planetary 
system could explain the abundance
pattern of the solar twins with respect to the Sun. \citet{cha10}
argues that just adding 4~\Mea of Earth-like material to
the current solar photosphere is enough to sufficiently increase 
the amount of refractory elements but not to reproduce the abundance 
pattern of solar twins in \citet{mel09}. 
To approximately recover that pattern one needs 
to add 4~\Mea of a equal mixture of Earth-like and CM 
chondrite material, provided that the fractionation is limited to
the current size of the convection zone, which comprises $\sim 2.5$\% 
of the Sun's mass. Thus, this augmented solar abundance pattern
roughly matches the mean abundance of those solar twins both for
volatile and refractory elements, suggesting that the Sun could have
accreted refractory-poor material from both the terrestrial-planet and
the asteroid-belt regions \citep{cha10}.  

One of the problems of this simple model is probably the size of the 
convection zone at the time of the formation of terrestrial planets.
This issue has already been discussed in \citet{mel09} where they 
point out that protoplanetary disks have typical lifetimes 
$\lesssim 10$~Myr \citep[see e.g.][]{sic06,mor12}, but at that time, 
the mass fraction of the convection zone of a solar-type star is 
about 40~\%~\citep{dan94}, and it goes down to its current size 
likely after 30~Myr. However, recently \citet{bar10} have shown
that episodic accretion during pre-main-sequence evolution of
solar-mass stars could accelerate the evolution of the convective
zone until reaching its current size. This effect is more evident
 as the initial mass of the protostars, 
before the onset of the accretion episodes, decreases. 
Thus, only in the case of initial mass of the protostars before the
onset of the accretion bursts as low as 10~$M_{\rm Jup}$ 
(with accretion rates of 0.5~$M_{\rm Jup}$~yr$^{-1}$
during episodes lasting for 100~yr) would the convection zone reach 
the current size in about 6-8~Myr, and therefore be consistent with 
the lifetime of the protoplanetary disk in the above scenario.
Different efficiencies of episodic accretion may also affect the 
final atmospheric abundances of the star, as can the composition
of the accreted material.

On the other hand, 
the accretion of small planets containing ``heavy'' refractory 
elements, which may happen, for instance, during inward migration 
of massive planets (as mentioned in 
section~\ref{secsa}), may also affect the composition of the 
convective zone of these stars depending on the relative strength 
of thermohaline mixing/convection caused by unstable $\mu$-gradient, 
external turbulence (as for instance rotation-induced
mixing), and radiative 
levitation~\citep[see][and references therein]{vau12}.
The remaining
question may be how much this affects the accretion of volatiles
elements such as C, N, and O, during accretion of the 
dust-depleted gas 
needed to explain the solar abundances in the above scenario.

Giant planets of the solar system also contain 
a significant number of refractory elements, 
$\sim 50-100$~\Mea~\citep{gui05}, but they do not appear 
to be enriched in rock-forming elements compared to 
ices~\citep{owe99}. Thus, it is difficult to estimate how much
giant-planet formation can affect the photospheric abundances of 
refractory elements in solar-type 
stars~\citep[see the discussion in][for more details]{cha10}. 

\citet{cha10} elaborates a simple model that quantifies the amount 
of rocky material in a stars' planetary system from the slopes of 
the trends described by the abundance ratios of a sample of solar 
analogs versus the condensation temperature, $T_C$. They use the
slopes derived by \citet{ram09} to tentatively identify stars
harboring terrestrial planets, according to their model.
We applied this model to our sample of solar analogs, but
we use the slopes determined with the average abundances in bins of
$\Delta T_C= 150$~K as described in Section~\ref{secse} and displayed 
in Fig.~\ref{fslpsa}. We use both
equations 1 and 2 in \citet{cha10} to estimate the mass of the 
accreted fractionated material, $M_{\rm frac}$, which is mixed with
material in the stellar convective zone, $M_{\rm CZ}$, and the mass 
of rocky material, $M_{\rm rock}$. We assume the mass of convective
zone to be constant over time and equal to 0.02~\Msun for all solar
analogs~\citep{pin01}. We also adopt the whole interval of 
condensation temperature 40--1750~K. We choose the value of 
$-0.35 \times 10^{-4}$~dex~K$^{-1}$ as minimum slope, 
which corresponds to the parameter $S_{\rm max}$ in \citet{cha10}, 
since most of the points are above this value.

In Fig.~\ref{fslpsat}, we depict
the same plot as Fig.~\ref{fslpsa}, but highlight those stars
hosting super-Earth-like, with minimum masses, $m_p \sin i$, 
between 5 and 14~\Meao, and Neptune-like, with 
$m_p \sin i$~14--30~\Meao, planets. In the middle 
panel of Fig.~\ref{fslpsat} we also display lines corresponding to
the amount of rocky material that is expected to be missed from 
the atmosphere of the stars according to their slopes. These lines
should track the amount of mass that is in terrestrial planets,
maybe also super-Earth-like planets, as well as possible asteroid
belts. 
This plot is similar to Figure 3 in \citet{cha10}, but the values 
used by \citet{cha10} are greater mainly because we 
subtract the chemical evolution effects from abundance trends of 
the stars, and \citet{ram09} did not. 
In addition, in this plot we distinguish among stars with and 
without known planets and with different minimum masses,
which are super-Earth-like, Neptune-like, and giant planets. 
We note that
we assume that at least super-Earth-like planets should contribute to
the amount of missing rocky material from the stellar atmospheres.
The position of the zero point depends on the adopted minimum slope,
so we may only consider these lines as a tentative indication of
the possible amount of terrestrial planets in the planetary systems of
these solar analogs. This is probably a weak point of this simple
model, which may only prevent us from drawing strong conclusions from
this comparison. However, by using this value we are able to roughly
reproduced the amount of rocky material in the solar planetary system,
$M_{\rm rock} \sim 4$~\Meao, which gives a flat slope at [Fe/H]$=0$.

Figure~\ref{fslpsat} reveals that most of the stars
may host certain amounts (more than 3~\Meao) of terrestrial planets. 
In particular, between 3~\Mea and 10~\Mea there are many stars with 
already detected super-Earth-like planets, but there is no clear 
correlation between the slope values and amount of missing rocky 
material. 
Stars with known super-Earth-like planets with similar mass have 
in some cases very different predicted missing masses of rock.
Thus, for instance, the star HD~45184 with only one super-Earth-like 
planet at $\sim 13$~\Mea exhibits a negative slope but the predicted
amount of rocky material is $\sim 2$~\Meao. The slope of the 
star HD~1461, which has two super-Earth-like planets accounting for 
a total mass of $\sim 14$~\Meao, yields a mass of rock of 
$\sim 6$~\Meao. The two stars with nearly flat slopes at about
$+0.05\times10^{-4}$~dex~K$^{-1}$, HD~31527 and HD~160691, and with 
low-mass planets at about $\sim 11$~\Meao, have however very different
predicted rocky masses at $\sim 3$ and $\sim 10$~\Meao, respectively.
This may mean that super-Earth-like planets are not traces of rocky
missing material but only Earth-like planets are, or that the 
assumption that the slope of abundance trends versus condensation 
temperature are not related to the amount of terrestrial planets in 
the planetary systems of solar analogs. 

In Fig.~\ref{fslpsat} we also depict stars with Neptune-like planets,
which may also have rocky cores covered by volatile-rich atmospheres. 
Similarly we do not really know if super-Earth-like planets have 
an atmosphere or a crust rich in volatiles.
Two stars with $\sim 17$~\Meao, have however significantly 
different predicted missing mass of rock from the stars' atmosphere 
at roughly 7~\Mea and 14~\Meao. The size of the cores of these two planets
are unknown, but if we assume that the cores are similar, then the
star at higher metallicity may also harbor some undetected 
Earth-like planets. However, again we cannot prove this is the
case but we also cannot demonstrate that the reasoning in
\citet{mel09} and \citet{ram09} is correct. 

\subsection{Conclusions}

We have explored the HARPS database to search for main-sequence 
stars with hotter effective temperature than the Sun and selected 
a sample in the \teff range 5950--6400~K of 32 ``single'' stars 
and 29 stars hosting planets with high-resolution 
($\lambda/\delta\lambda \gtrsim 85,000$) and 
very high signal-to-noise (S/N~$\gtrsim 250$ and S/N~$\sim 800$ on 
average) HARPS, UVES, and UES spectroscopic data. 
We tried to evaluate the possible connection between the
abundance pattern versus the condensation temperature, $T_C$, 
and the presence of terrestrial planets. 

We performed a detailed chemical abundance analysis of this sample 
of ``hot'' analogs and 
investigated possible trends of mean abundance ratios, 
$\Delta {\rm [X/Fe]_{\rm SUN-STARS}}$, versus $T_C$, after removing 
the Galactic chemical evolution effects.
We find that stars both with and without planets show similar
mean abundance patterns with slightly negative slopes.
We also analyzed a subsample of metal-rich ``hot'' analogs in the
narrow metallicity range +0.04--+0.19 and also found negative slopes
for stars both with and without planets, and almost equal mean
abundance patterns when restricting the sample to very high quality
spectra (S/N $> 550$).

We also revisited the sample of solar analogs in \citet{gon10} 
with the updated number of planet hosts, 62 ``single'' stars 
and 33 stars with planets. 
We investigated the mean abundance pattern of solar
analogs harboring planets with the most massive planets at different 
orbital periods and with the least massive planet being a
super-Earth-like, a Neptune-like, and a Jupiter-like planet. We
speculated within the scenario in which the presence of rocky planets 
affect the volatile-to-refractory abundance ratios in the atmosphere 
of these planet-host stars. 
In each case the slope value reveals expected value 
according to this scenario, assuming that the bulk chemical 
composition of super-Earth planets is closer to the Earth's 
composition than that of Neptunian and Jovian planets.  

This sample of solar analogs and of ``hot'' analogs have a 
subsample of eight and two 
stars harboring super-Earth-like planets. We computed the average 
abundance ratios $\Delta {\rm [X/Fe]_{\rm SUN-STARS}}$ versus 
condensation, $T_C$, in bins of $\Delta T_C = 150$~K, and 
individually fit these abundances to get the slope of the average 
abundance pattern of each star.
We find that only four stars show positive abundance trends,
qualitatively in agreement with the expected value according the
reasoning in \citet{mel09} and \citet{ram09}. 
However, three stars show nearly 
flat slopes, and three stars exhibit unexpected negative trends since
they harbor super-Earth-like planets. 

Finally, we derived the slopes of the trend 
$\Delta {\rm [X/Fe]_{\rm SUN-STARS}}$ versus $T_C$ for each star 
of the whole sample of solar analogs and compared them with a simple
model of the amount of rocky material missed in the stars'
atmospheres. Several solar analogs containing super-Earth-like 
planets of typically $m_p \sin i \sim 11$~\Meao, 
with different slopes, provide very inconsistent predictions of 
the missing amount of rocks from 2 to 15~\Meao. 
This might mean that only Earth-like planets could track the amount
of missing rocky material in the atmospheres of solar analogs.
Therefore, this comparison cannot provide clear evidence that
in the abundance pattern of a solar-type star there are hints of 
terrestrial planets in the planetary system of each star.

\begin{acknowledgements}
J.I.G.H. and G.I. acknowledge financial support from the 
Spanish Ministry project MICINN AYA2011-29060 and J.I.G.H. 
also from the Spanish Ministry of Science and Innovation 
(MICINN) under the 2009 Juan de la Cierva Program. 
This work was also supported by the European Research
Council/European Community under the FP7/EC through a Starting 
Grant agreement number 239953, as well as by Funda\c{c}\~ao 
para a Ci\^encia e Tecnologia (FCT) in the form of grant 
reference PTDC/CTE-AST/098528/2008. 
N.C.S. would further like to thank FCT through program 
Ci\^encia 2007 funded by FCT/MCTES (Portugal) and POPH/FSE (EC). 
V.Zh.A. and S.G.S. are supported by grants SFRH/BPD/70574/2010 
and SFRH/BPD/47611/2008 from FCT (Portugal), respectively.
E.D.M is supported by grant SFRH/BPD/76606/2011 from FCT
(Portugal).
This research made use of the SIMBAD database
operated at the CDS, Strasbourg, France. 
This work also made use of the IRAF facility and the 
Encyclopaedia of extrasolar planets.
\end{acknowledgements}

\bibliographystyle{aa}
\bibliography{planet}

\clearpage

\Online

\begin{appendix}

\section{Chemical abundance ratios [X/Fe]}
In this appendix we provide all the tables containing the element 
abundance ratios [X/Fe] of 32 F-, G-type stars without planets 
and 29 F-, G-type stars hosting planets that are all available online. 

\begin{table*}
\centering
\caption{Abundance ratios [X/Fe] of F-, G-type analogs with planets\label{tabonline}}
\begin{tabular}{lrrrrrr}
\noalign{\smallskip}
\noalign{\smallskip}
\noalign{\smallskip}
\hline
\hline
\noalign{\medskip} 
HD       &     [C/Fe]      &     [O/Fe]      &     [S/Fe]      &     [Na/Fe]      &     [Mg/Fe]      &     [Al/Fe]      \\
\noalign{\medskip} 
\hline
\hline
\noalign{\medskip} 
   10647 &        --        & $ 0.112\pm0.070$ & $-0.058\pm0.042$ & $-0.078\pm0.057$ & $-0.063\pm0.064$ & $-0.148\pm0.014$ \\
  108147 & $-0.131\pm0.028$ & $-0.181\pm0.070$ & $-0.136\pm0.007$ & $-0.051\pm0.057$ & $-0.041\pm0.057$ & $-0.116\pm0.007$ \\
  117618 & $-0.060\pm0.049$ & $-0.003\pm0.070$ & $-0.008\pm0.049$ & $ 0.057\pm0.042$ & $ 0.002\pm0.021$ & $-0.018\pm0.021$ \\
  121504 & $-0.112\pm0.042$ & $ 0.014\pm0.070$ & $-0.061\pm0.120$ & $-0.126\pm0.085$ & $-0.071\pm0.049$ & $-0.051\pm0.035$ \\
  134060 & $-0.070\pm0.015$ & $-0.027\pm0.070$ & $-0.052\pm0.021$ & $ 0.023\pm0.057$ & $-0.027\pm0.042$ & $-0.037\pm0.014$ \\
  169830 & $-0.018\pm0.068$ & $ 0.068\pm0.070$ & $-0.027\pm0.092$ & $ 0.048\pm0.057$ & $-0.037\pm0.021$ & $-0.097\pm0.021$ \\
   17051 &        --        & $-0.061\pm0.070$ & $-0.066\pm0.007$ & $ 0.019\pm0.070$ & $-0.011\pm0.113$ & $-0.111\pm0.070$ \\
  179949 & $-0.017\pm0.127$ & $ 0.000\pm0.070$ & $-0.070\pm0.028$ & $ 0.020\pm0.014$ & $-0.120\pm0.028$ & $-0.260\pm0.028$ \\
   19994 & $ 0.018\pm0.068$ & $ 0.061\pm0.070$ & $ 0.061\pm0.070$ & $ 0.136\pm0.021$ &        --        & $-0.049\pm0.070$ \\
  208487 & $-0.064\pm0.030$ & $ 0.016\pm0.070$ & $-0.024\pm0.070$ & $ 0.016\pm0.042$ & $ 0.031\pm0.106$ & $-0.064\pm0.028$ \\
  212301 & $-0.094\pm0.028$ & $ 0.086\pm0.070$ & $-0.114\pm0.057$ & $-0.004\pm0.071$ & $-0.054\pm0.057$ & $-0.219\pm0.021$ \\
  216435 &        --        & $-0.044\pm0.070$ & $-0.024\pm0.070$ & $ 0.111\pm0.021$ & $-0.014\pm0.014$ & $ 0.006\pm0.042$ \\
  221287 &        --        & $ 0.037\pm0.070$ & $-0.178\pm0.021$ & $-0.078\pm0.021$ & $-0.113\pm0.014$ & $-0.173\pm0.070$ \\
   23079 & $-0.001\pm0.045$ & $ 0.103\pm0.070$ & $-0.047\pm0.042$ & $-0.022\pm0.064$ & $ 0.008\pm0.021$ & $-0.027\pm0.028$ \\
   39091 &        --        & $ 0.029\pm0.070$ & $-0.071\pm0.028$ & $ 0.039\pm0.057$ & $ 0.029\pm0.042$ & $-0.046\pm0.035$ \\
    7449 & $-0.039\pm0.035$ & $ 0.164\pm0.070$ & $-0.055\pm0.042$ & $-0.035\pm0.057$ &        --        & $-0.090\pm0.007$ \\
   75289 & $-0.168\pm0.025$ & $-0.165\pm0.070$ & $-0.155\pm0.127$ & $-0.050\pm0.035$ & $ 0.005\pm0.042$ & $-0.060\pm0.007$ \\
   82943 & $-0.069\pm0.021$ & $-0.036\pm0.070$ & $-0.056\pm0.085$ & $ 0.094\pm0.070$ & $-0.071\pm0.120$ & $-0.031\pm0.021$ \\
   93385 & $-0.045\pm0.026$ & $ 0.055\pm0.070$ & $-0.050\pm0.064$ & $ 0.040\pm0.049$ & $-0.025\pm0.028$ & $-0.015\pm0.014$ \\
  209458 & $ 0.011\pm0.070$ & $-0.099\pm0.070$ & $-0.084\pm0.064$ & $-0.044\pm0.064$ &        --        & $-0.089\pm0.014$ \\
    2039 & $-0.073\pm0.106$ & $-0.143\pm0.070$ & $-0.083\pm0.042$ & $ 0.107\pm0.028$ & $-0.123\pm0.070$ & $-0.003\pm0.028$ \\
  213240 & $-0.075\pm0.025$ & $ 0.061\pm0.070$ & $-0.122\pm0.029$ & $ 0.071\pm0.042$ & $ 0.041\pm0.042$ & $ 0.021\pm0.028$ \\
   23596 & $ 0.019\pm0.076$ & $-0.008\pm0.070$ & $-0.043\pm0.035$ & $ 0.162\pm0.071$ & $-0.018\pm0.014$ & $-0.008\pm0.014$ \\
   50554 & $ 0.033\pm0.101$ & $-0.094\pm0.070$ & $-0.129\pm0.035$ & $-0.064\pm0.042$ & $-0.034\pm0.042$ & $-0.084\pm0.014$ \\
   52265 & $-0.060\pm0.047$ & $ 0.053\pm0.070$ & $ 0.013\pm0.057$ & $ 0.078\pm0.035$ &        --        & $-0.057\pm0.028$ \\
   72659 & $ 0.020\pm0.040$ & $ 0.030\pm0.070$ & $-0.050\pm0.085$ & $ 0.060\pm0.042$ & $ 0.030\pm0.070$ & $ 0.010\pm0.014$ \\
   74156 & $-0.037\pm0.025$ & $ 0.006\pm0.070$ & $-0.054\pm0.071$ & $ 0.076\pm0.057$ & $ 0.026\pm0.070$ & $-0.034\pm0.014$ \\
   89744 & $-0.078\pm0.014$ & $-0.088\pm0.070$ &        --        & $ 0.052\pm0.042$ & $ 0.142\pm0.070$ &        --        \\
    9826 & $-0.003\pm0.007$ & $-0.078\pm0.070$ &        --        & $ 0.052\pm0.042$ & $ 0.102\pm0.070$ & $ 0.037\pm0.148$ \\
\noalign{\medskip} 
\hline
\end{tabular}
\end{table*}

\clearpage

\begin{table*}
\centering
\caption{Abundance ratios [X/Fe] of F-, G-type analogs with planets\label{tabonline}}
\begin{tabular}{lrrrrrr}
\noalign{\smallskip}
\noalign{\smallskip}
\noalign{\smallskip}
\hline
\hline
\noalign{\medskip} 
HD       &     [Si/Fe]      &     [Ca/Fe]      &     [Sc/Fe]      &     [Ti/Fe]      &     [V/Fe]      &     [Cr/Fe]      \\
\noalign{\medskip} 
\hline
\hline
\noalign{\medskip} 
   10647 & $-0.031\pm0.016$ & $ 0.023\pm0.024$ & $-0.033\pm0.031$ & $-0.016\pm0.031$ & $-0.098\pm0.049$ & $-0.022\pm0.040$ \\
  108147 & $-0.010\pm0.019$ & $ 0.028\pm0.021$ & $-0.028\pm0.010$ & $-0.016\pm0.033$ & $-0.054\pm0.041$ & $-0.029\pm0.025$ \\
  117618 & $ 0.004\pm0.018$ & $ 0.024\pm0.032$ & $ 0.029\pm0.030$ & $-0.019\pm0.016$ & $-0.018\pm0.006$ & $-0.006\pm0.017$ \\
  121504 & $-0.016\pm0.015$ & $ 0.021\pm0.024$ & $ 0.000\pm0.034$ & $-0.014\pm0.017$ & $-0.050\pm0.015$ & $ 0.014\pm0.007$ \\
  134060 & $-0.006\pm0.025$ & $ 0.008\pm0.013$ & $ 0.039\pm0.032$ & $ 0.005\pm0.017$ & $-0.007\pm0.041$ & $-0.002\pm0.017$ \\
  169830 & $ 0.007\pm0.016$ & $ 0.014\pm0.048$ & $-0.004\pm0.027$ & $-0.014\pm0.034$ & $-0.064\pm0.030$ & $-0.021\pm0.023$ \\
   17051 & $-0.006\pm0.029$ & $ 0.017\pm0.031$ & $-0.004\pm0.074$ & $-0.003\pm0.050$ & $-0.021\pm0.084$ & $ 0.002\pm0.019$ \\
  179949 & $-0.013\pm0.040$ & $ 0.005\pm0.033$ & $-0.040\pm0.010$ & $-0.022\pm0.037$ & $-0.018\pm0.058$ & $-0.025\pm0.026$ \\
   19994 & $ 0.027\pm0.032$ & $-0.007\pm0.029$ & $ 0.076\pm0.019$ & $ 0.041\pm0.044$ & $-0.005\pm0.075$ & $-0.042\pm0.056$ \\
  208487 & $-0.004\pm0.009$ & $ 0.021\pm0.010$ & $ 0.004\pm0.022$ & $-0.004\pm0.024$ & $-0.062\pm0.040$ & $-0.018\pm0.019$ \\
  212301 & $-0.019\pm0.034$ & $ 0.014\pm0.016$ & $-0.014\pm0.048$ & $-0.016\pm0.034$ & $-0.006\pm0.041$ & $ 0.001\pm0.021$ \\
  216435 & $ 0.018\pm0.015$ & $ 0.019\pm0.029$ & $ 0.051\pm0.025$ & $-0.008\pm0.035$ & $ 0.012\pm0.044$ & $-0.018\pm0.027$ \\
  221287 & $-0.035\pm0.014$ & $ 0.044\pm0.032$ & $-0.028\pm0.017$ & $ 0.020\pm0.034$ & $-0.066\pm0.088$ & $-0.009\pm0.043$ \\
   23079 & $ 0.008\pm0.007$ & $ 0.051\pm0.010$ & $ 0.037\pm0.036$ & $ 0.029\pm0.021$ & $-0.064\pm0.040$ & $-0.002\pm0.020$ \\
   39091 & $ 0.001\pm0.013$ & $ 0.013\pm0.017$ & $ 0.019\pm0.042$ & $-0.018\pm0.026$ & $-0.038\pm0.025$ & $-0.002\pm0.013$ \\
    7449 & $ 0.001\pm0.007$ & $ 0.052\pm0.018$ & $-0.017\pm0.026$ & $ 0.020\pm0.025$ & $-0.054\pm0.045$ & $-0.000\pm0.026$ \\
   75289 & $-0.028\pm0.028$ & $ 0.045\pm0.030$ & $ 0.043\pm0.054$ & $ 0.010\pm0.024$ & $-0.018\pm0.045$ & $ 0.002\pm0.014$ \\
   82943 & $-0.006\pm0.018$ & $-0.036\pm0.062$ & $ 0.064\pm0.037$ & $-0.027\pm0.026$ & $-0.004\pm0.023$ & $ 0.000\pm0.017$ \\
   93385 & $ 0.008\pm0.013$ & $ 0.020\pm0.010$ & $ 0.028\pm0.021$ & $-0.006\pm0.023$ & $-0.035\pm0.042$ & $ 0.001\pm0.019$ \\
  209458 & $ 0.002\pm0.056$ & $ 0.041\pm0.055$ &        --        & $ 0.014\pm0.090$ & $-0.069\pm0.029$ & $ 0.024\pm0.133$ \\
    2039 & $-0.000\pm0.017$ & $ 0.014\pm0.036$ & $ 0.039\pm0.047$ & $-0.000\pm0.029$ & $ 0.047\pm0.007$ & $ 0.006\pm0.012$ \\
  213240 & $ 0.007\pm0.012$ & $ 0.008\pm0.032$ & $ 0.071\pm0.055$ & $ 0.008\pm0.023$ & $ 0.015\pm0.025$ & $-0.022\pm0.016$ \\
   23596 & $ 0.014\pm0.039$ & $-0.027\pm0.045$ & $ 0.105\pm0.022$ & $-0.025\pm0.034$ & $ 0.006\pm0.023$ & $-0.013\pm0.039$ \\
   50554 & $-0.009\pm0.049$ & $ 0.009\pm0.034$ & $ 0.049\pm0.006$ & $ 0.009\pm0.049$ & $ 0.008\pm0.058$ & $-0.025\pm0.044$ \\
   52265 & $ 0.015\pm0.020$ & $ 0.061\pm0.033$ & $ 0.028\pm0.033$ & $-0.027\pm0.021$ & $-0.031\pm0.023$ & $-0.015\pm0.033$ \\
   72659 & $ 0.021\pm0.012$ & $ 0.026\pm0.014$ & $ 0.050\pm0.026$ & $ 0.014\pm0.015$ & $-0.030\pm0.038$ & $-0.008\pm0.008$ \\
   74156 & $ 0.010\pm0.010$ & $ 0.030\pm0.038$ & $ 0.044\pm0.010$ & $ 0.000\pm0.029$ & $-0.017\pm0.027$ & $-0.034\pm0.019$ \\
   89744 & $ 0.015\pm0.034$ & $ 0.030\pm0.034$ & $ 0.035\pm0.043$ & $-0.021\pm0.052$ & $-0.036\pm0.067$ & $-0.038\pm0.028$ \\
    9826 & $ 0.032\pm0.031$ & $ 0.017\pm0.014$ & $ 0.080\pm0.024$ & $ 0.009\pm0.058$ & $ 0.010\pm0.077$ & $-0.043\pm0.044$ \\
\noalign{\medskip} 
\hline
\end{tabular}
\end{table*}

\clearpage

\begin{table*}
\centering
\caption{Abundance ratios [X/Fe] of F-, G-type analogs with planets\label{tabonline}}
\begin{tabular}{lrrrrrr}
\noalign{\smallskip}
\noalign{\smallskip}
\noalign{\smallskip}
\hline
\hline
\noalign{\medskip} 
HD       &     [Mn/Fe]      &     [Co/Fe]      &     [Ni/Fe]      &     [Cu/Fe]      &     [Zn/Fe]      &     [Sr/Fe]      \\
\noalign{\medskip} 
\hline
\hline
\noalign{\medskip} 
   10647 & $-0.046\pm0.083$ & $-0.122\pm0.037$ & $-0.064\pm0.024$ & $-0.048\pm0.070$ & $-0.078\pm0.028$ & $ 0.092\pm0.070$ \\
  108147 & $-0.048\pm0.033$ & $-0.088\pm0.024$ & $-0.036\pm0.030$ & $-0.061\pm0.070$ & $-0.091\pm0.042$ & $ 0.069\pm0.070$ \\
  117618 & $-0.006\pm0.026$ & $-0.027\pm0.009$ & $ 0.003\pm0.015$ & $ 0.017\pm0.070$ & $ 0.007\pm0.014$ & $-0.043\pm0.070$ \\
  121504 & $-0.048\pm0.021$ & $-0.049\pm0.005$ & $-0.018\pm0.016$ & $-0.026\pm0.070$ & $-0.016\pm0.057$ & $ 0.024\pm0.070$ \\
  134060 & $-0.005\pm0.028$ & $-0.007\pm0.006$ & $ 0.008\pm0.022$ & $-0.007\pm0.070$ & $ 0.003\pm0.085$ & $-0.017\pm0.070$ \\
  169830 & $-0.052\pm0.054$ & $-0.035\pm0.024$ & $-0.011\pm0.023$ & $ 0.018\pm0.070$ & $-0.077\pm0.021$ & $-0.002\pm0.070$ \\
   17051 & $ 0.006\pm0.046$ & $-0.063\pm0.027$ & $-0.013\pm0.027$ & $-0.051\pm0.070$ & $-0.051\pm0.028$ & $ 0.039\pm0.070$ \\
  179949 & $ 0.028\pm0.021$ & $-0.066\pm0.029$ & $-0.025\pm0.028$ & $-0.050\pm0.070$ & $-0.110\pm0.042$ & $ 0.000\pm0.070$ \\
   19994 & $ 0.069\pm0.017$ & $ 0.006\pm0.051$ & $ 0.037\pm0.028$ & $ 0.121\pm0.070$ & $-0.059\pm0.070$ & $-0.019\pm0.070$ \\
  208487 & $-0.074\pm0.067$ & $-0.046\pm0.029$ & $-0.025\pm0.019$ & $-0.004\pm0.070$ & $-0.024\pm0.057$ & $-0.004\pm0.070$ \\
  212301 & $ 0.003\pm0.094$ & $-0.036\pm0.056$ & $-0.009\pm0.030$ & $ 0.006\pm0.070$ & $-0.064\pm0.042$ & $-0.034\pm0.070$ \\
  216435 & $ 0.042\pm0.073$ & $ 0.016\pm0.015$ & $ 0.028\pm0.026$ & $ 0.026\pm0.070$ & $ 0.011\pm0.120$ & $-0.104\pm0.070$ \\
  221287 & $-0.260\pm0.103$ & $-0.116\pm0.034$ & $-0.068\pm0.032$ & $-0.013\pm0.070$ & $-0.093\pm0.070$ & $ 0.107\pm0.070$ \\
   23079 & $-0.070\pm0.070$ & $-0.039\pm0.024$ & $-0.031\pm0.014$ & $-0.007\pm0.070$ & $-0.037\pm0.014$ & $-0.017\pm0.070$ \\
   39091 & $-0.009\pm0.031$ & $-0.025\pm0.005$ & $-0.003\pm0.010$ & $-0.001\pm0.070$ & $ 0.019\pm0.042$ & $-0.061\pm0.070$ \\
    7449 & $-0.063\pm0.079$ & $-0.085\pm0.017$ & $-0.044\pm0.012$ & $-0.035\pm0.070$ & $-0.055\pm0.014$ & $ 0.005\pm0.070$ \\
   75289 & $-0.060\pm0.057$ & $-0.039\pm0.022$ & $-0.015\pm0.026$ & $-0.065\pm0.070$ & $-0.145\pm0.057$ & $ 0.015\pm0.070$ \\
   82943 & $ 0.072\pm0.019$ & $ 0.010\pm0.025$ & $ 0.029\pm0.019$ & $ 0.004\pm0.070$ & $-0.006\pm0.042$ & $-0.086\pm0.070$ \\
   93385 & $-0.049\pm0.045$ & $-0.033\pm0.017$ & $-0.002\pm0.012$ & $-0.005\pm0.070$ & $ 0.000\pm0.007$ & $-0.035\pm0.070$ \\
  209458 & $-0.146\pm0.062$ & $-0.051\pm0.133$ & $-0.032\pm0.110$ &        --        & $-0.139\pm0.070$ & $-0.009\pm0.070$ \\
    2039 & $ 0.097\pm0.070$ & $ 0.033\pm0.023$ & $ 0.039\pm0.038$ & $-0.033\pm0.070$ & $-0.053\pm0.070$ &        --        \\
  213240 & $-0.014\pm0.029$ & $ 0.016\pm0.012$ & $ 0.013\pm0.020$ & $-0.009\pm0.070$ & $ 0.028\pm0.040$ & $-0.069\pm0.070$ \\
   23596 & $ 0.137\pm0.031$ & $ 0.019\pm0.030$ & $ 0.042\pm0.029$ &        --        & $ 0.062\pm0.113$ & $-0.108\pm0.070$ \\
   50554 & $-0.076\pm0.102$ & $-0.057\pm0.025$ & $-0.035\pm0.037$ &        --        & $-0.069\pm0.134$ & $ 0.216\pm0.070$ \\
   52265 & $-0.015\pm0.067$ & $-0.040\pm0.031$ & $ 0.006\pm0.026$ & $-0.017\pm0.070$ & $-0.007\pm0.017$ & $-0.007\pm0.070$ \\
   72659 & $-0.110\pm0.071$ & $ 0.005\pm0.005$ & $-0.007\pm0.020$ & $ 0.060\pm0.070$ & $-0.020\pm0.070$ &        --        \\
   74156 & $-0.059\pm0.092$ & $-0.004\pm0.012$ & $ 0.016\pm0.028$ & $ 0.046\pm0.070$ & $-0.034\pm0.070$ &        --        \\
   89744 & $ 0.082\pm0.099$ & $-0.074\pm0.047$ & $-0.038\pm0.044$ &        --        &        --        &        --        \\
    9826 &        --        & $-0.031\pm0.032$ & $-0.017\pm0.039$ &        --        & $-0.008\pm0.070$ &        --        \\
\noalign{\medskip} 
\hline
\end{tabular}
\end{table*}

\clearpage

\begin{table*}
\centering
\caption{Abundance ratios [X/Fe] of F-, G-type analogs with planets\label{tabonline}}
\begin{tabular}{lrrrrrr}
\noalign{\smallskip}
\noalign{\smallskip}
\noalign{\smallskip}
\hline
\hline
\noalign{\medskip} 
HD       &     [Y/Fe]      &     [Zr/Fe]      &     [Ba/Fe]      &     [Ce/Fe]      &     [Nd/Fe]      &     [Eu/Fe]      \\
\noalign{\medskip} 
\hline
\hline
\noalign{\medskip} 
   10647 & $ 0.162\pm0.026$ & $ 0.172\pm0.070$ & $ 0.262\pm0.042$ & $ 0.135\pm0.074$ & $ 0.032\pm0.070$ & $ 0.062\pm0.070$ \\
  108147 & $ 0.124\pm0.049$ & $ 0.119\pm0.070$ & $ 0.124\pm0.064$ & $ 0.042\pm0.006$ & $-0.021\pm0.070$ & $ 0.169\pm0.070$ \\
  117618 & $ 0.047\pm0.020$ & $ 0.087\pm0.070$ & $ 0.057\pm0.070$ & $-0.070\pm0.031$ & $-0.003\pm0.070$ & $ 0.027\pm0.070$ \\
  121504 & $ 0.098\pm0.035$ & $ 0.114\pm0.070$ & $ 0.149\pm0.021$ & $ 0.038\pm0.015$ & $-0.026\pm0.070$ & $ 0.074\pm0.070$ \\
  134060 & $ 0.033\pm0.040$ & $ 0.053\pm0.070$ & $ 0.038\pm0.021$ & $ 0.003\pm0.017$ & $-0.137\pm0.070$ & $-0.027\pm0.070$ \\
  169830 & $ 0.068\pm0.044$ & $ 0.058\pm0.070$ & $ 0.033\pm0.021$ & $-0.055\pm0.067$ & $-0.082\pm0.070$ & $-0.052\pm0.070$ \\
   17051 & $ 0.092\pm0.055$ & $ 0.089\pm0.070$ & $ 0.128\pm0.099$ & $ 0.029\pm0.026$ & $-0.161\pm0.070$ & $ 0.039\pm0.070$ \\
  179949 & $ 0.050\pm0.028$ & $ 0.050\pm0.070$ & $ 0.020\pm0.028$ & $-0.027\pm0.012$ & $-0.120\pm0.070$ & $ 0.060\pm0.070$ \\
   19994 & $ 0.088\pm0.090$ & $ 0.141\pm0.070$ & $-0.124\pm0.007$ & $-0.022\pm0.057$ & $-0.079\pm0.070$ & $ 0.201\pm0.070$ \\
  208487 & $ 0.069\pm0.050$ & $ 0.086\pm0.070$ & $ 0.081\pm0.021$ & $ 0.049\pm0.021$ & $ 0.006\pm0.070$ & $ 0.026\pm0.070$ \\
  212301 & $ 0.062\pm0.025$ & $ 0.086\pm0.070$ & $-0.019\pm0.021$ &        --        & $-0.084\pm0.070$ & $ 0.025\pm0.070$ \\
  216435 & $ 0.050\pm0.059$ & $-0.024\pm0.070$ & $-0.009\pm0.021$ & $-0.030\pm0.035$ & $-0.124\pm0.070$ & $-0.084\pm0.070$ \\
  221287 & $ 0.174\pm0.021$ & $ 0.207\pm0.070$ & $ 0.297\pm0.085$ & $ 0.107\pm0.123$ & $ 0.067\pm0.070$ & $-0.013\pm0.070$ \\
   23079 & $ 0.089\pm0.029$ & $ 0.123\pm0.070$ & $ 0.113\pm0.070$ & $ 0.049\pm0.051$ & $ 0.013\pm0.070$ & $ 0.073\pm0.070$ \\
   39091 & $ 0.035\pm0.031$ & $-0.011\pm0.070$ & $ 0.019\pm0.070$ & $-0.011\pm0.036$ & $-0.111\pm0.070$ & $-0.051\pm0.070$ \\
    7449 & $ 0.101\pm0.029$ & $ 0.105\pm0.070$ & $ 0.185\pm0.070$ & $ 0.095\pm0.052$ & $ 0.025\pm0.070$ & $ 0.125\pm0.070$ \\
   75289 & $ 0.099\pm0.042$ & $ 0.095\pm0.070$ & $ 0.075\pm0.070$ & $ 0.009\pm0.023$ & $-0.045\pm0.070$ & $-0.055\pm0.070$ \\
   82943 & $-0.016\pm0.040$ & $ 0.014\pm0.070$ & $-0.051\pm0.049$ & $-0.039\pm0.064$ & $-0.116\pm0.070$ & $-0.056\pm0.070$ \\
   93385 & $ 0.042\pm0.021$ & $ 0.065\pm0.070$ & $ 0.080\pm0.021$ & $ 0.002\pm0.035$ & $-0.065\pm0.070$ & $ 0.025\pm0.070$ \\
  209458 &        --        &        --        & $ 0.211\pm0.085$ & $ 0.041\pm0.072$ &        --        & $ 0.081\pm0.070$ \\
    2039 & $ 0.001\pm0.015$ & $-0.043\pm0.070$ & $-0.108\pm0.021$ &        --        &        --        & $-0.073\pm0.070$ \\
  213240 & $ 0.005\pm0.042$ & $ 0.001\pm0.070$ & $-0.019\pm0.028$ & $-0.005\pm0.006$ & $-0.119\pm0.070$ & $ 0.011\pm0.070$ \\
   23596 & $ 0.042\pm0.035$ & $-0.028\pm0.070$ & $-0.108\pm0.070$ & $-0.118\pm0.141$ & $-0.208\pm0.070$ & $-0.128\pm0.070$ \\
   50554 & $ 0.163\pm0.045$ & $ 0.136\pm0.070$ & $ 0.136\pm0.085$ & $ 0.321\pm0.092$ & $ 0.026\pm0.070$ & $ 0.026\pm0.070$ \\
   52265 & $ 0.050\pm0.047$ & $ 0.063\pm0.070$ & $-0.012\pm0.007$ & $-0.071\pm0.025$ & $-0.107\pm0.070$ & $ 0.013\pm0.070$ \\
   72659 & $-0.030\pm0.044$ & $ 0.020\pm0.070$ & $-0.005\pm0.007$ &        --        & $-0.080\pm0.070$ & $-0.030\pm0.070$ \\
   74156 & $ 0.050\pm0.058$ & $-0.004\pm0.070$ & $-0.009\pm0.021$ &        --        & $-0.024\pm0.070$ & $ 0.056\pm0.070$ \\
   89744 &        --        & $ 0.142\pm0.070$ & $ 0.022\pm0.070$ &        --        &        --        & $ 0.182\pm0.070$ \\
    9826 &        --        & $ 0.232\pm0.070$ & $ 0.022\pm0.028$ &        --        &        --        & $ 0.102\pm0.070$ \\
\noalign{\medskip} 
\hline
\end{tabular}
\end{table*}

\clearpage

\begin{table*}
\centering
\caption{Abundance ratios [X/Fe] of F-, G-type analogs without known planets\label{tabonline}}
\begin{tabular}{lrrrrrr}
\noalign{\smallskip}
\noalign{\smallskip}
\noalign{\smallskip}
\hline
\hline
\noalign{\medskip} 
HD       &     [C/Fe]      &     [O/Fe]      &     [S/Fe]      &     [Na/Fe]      &     [Mg/Fe]      &     [Al/Fe]      \\
\noalign{\medskip} 
\hline
\hline
\noalign{\medskip} 
   11226 & $ 0.060\pm0.042$ & $ 0.054\pm0.070$ & $ 0.083\pm0.070$ & $ 0.073\pm0.042$ &        --        & $-0.042\pm0.021$ \\
  119638 & $ 0.016\pm0.044$ & $ 0.116\pm0.070$ & $ 0.006\pm0.057$ & $ 0.021\pm0.064$ & $-0.029\pm0.035$ & $-0.054\pm0.028$ \\
  122862 & $ 0.050\pm0.035$ & $ 0.094\pm0.070$ & $-0.016\pm0.042$ & $ 0.044\pm0.042$ & $-0.001\pm0.021$ & $-0.021\pm0.035$ \\
  125881 & $-0.037\pm0.070$ & $-0.037\pm0.070$ & $-0.032\pm0.049$ & $-0.012\pm0.035$ &        --        & $-0.072\pm0.021$ \\
    1388 & $-0.068\pm0.020$ & $ 0.032\pm0.070$ & $-0.043\pm0.078$ & $ 0.002\pm0.042$ & $-0.068\pm0.099$ & $-0.033\pm0.007$ \\
  145666 & $-0.054\pm0.071$ & $-0.117\pm0.070$ & $-0.127\pm0.070$ & $-0.042\pm0.035$ & $-0.037\pm0.014$ & $-0.087\pm0.028$ \\
  157338 & $-0.011\pm0.035$ & $ 0.094\pm0.070$ & $-0.036\pm0.028$ & $ 0.009\pm0.064$ & $-0.016\pm0.070$ & $-0.056\pm0.028$ \\
    1581 & $ 0.004\pm0.060$ & $ 0.067\pm0.070$ & $-0.038\pm0.007$ & $ 0.022\pm0.049$ & $ 0.042\pm0.035$ & $-0.018\pm0.035$ \\
  168871 &        --        & $ 0.027\pm0.070$ & $ 0.007\pm0.070$ & $ 0.042\pm0.035$ & $ 0.052\pm0.064$ & $-0.003\pm0.014$ \\
  171990 & $ 0.034\pm0.023$ & $ 0.118\pm0.070$ & $ 0.068\pm0.070$ & $ 0.093\pm0.049$ & $ 0.038\pm0.028$ & $-0.017\pm0.007$ \\
  193193 & $ 0.002\pm0.029$ & $ 0.205\pm0.070$ & $-0.020\pm0.064$ & $ 0.055\pm0.042$ & $-0.010\pm0.035$ & $-0.005\pm0.028$ \\
  196800 & $-0.097\pm0.060$ & $ 0.063\pm0.070$ & $-0.007\pm0.057$ & $ 0.088\pm0.035$ & $-0.042\pm0.064$ & $-0.007\pm0.070$ \\
  199960 & $-0.021\pm0.014$ & $ 0.069\pm0.070$ &        --        & $ 0.059\pm0.113$ & $-0.056\pm0.092$ & $-0.021\pm0.057$ \\
  204385 & $ 0.005\pm0.090$ & $ 0.032\pm0.070$ & $-0.053\pm0.035$ & $ 0.052\pm0.042$ & $-0.003\pm0.007$ & $-0.018\pm0.028$ \\
  221356 &        --        & $ 0.115\pm0.070$ & $-0.175\pm0.028$ & $-0.070\pm0.078$ & $ 0.025\pm0.071$ & $-0.045\pm0.028$ \\
   31822 & $-0.061\pm0.021$ & $ 0.096\pm0.070$ & $-0.114\pm0.070$ & $-0.059\pm0.049$ & $-0.034\pm0.070$ & $-0.089\pm0.035$ \\
   36379 & $ 0.070\pm0.026$ & $ 0.110\pm0.070$ & $ 0.041\pm0.070$ & $ 0.055\pm0.049$ & $ 0.050\pm0.042$ & $-0.020\pm0.014$ \\
    3823 & $ 0.110\pm0.085$ & $ 0.227\pm0.070$ & $ 0.012\pm0.064$ & $ 0.032\pm0.049$ & $ 0.087\pm0.071$ & $-0.003\pm0.042$ \\
   38973 & $ 0.002\pm0.025$ & $-0.025\pm0.070$ & $-0.025\pm0.057$ & $ 0.025\pm0.028$ & $-0.030\pm0.035$ & $-0.030\pm0.021$ \\
   44120 & $-0.030\pm0.038$ & $-0.006\pm0.070$ & $-0.006\pm0.070$ & $ 0.069\pm0.049$ & $-0.036\pm0.042$ & $-0.016\pm0.028$ \\
   44447 & $ 0.060\pm0.061$ & $ 0.190\pm0.070$ &        --        & $ 0.030\pm0.042$ & $ 0.090\pm0.085$ & $-0.020\pm0.014$ \\
    6735 & $-0.029\pm0.042$ & $ 0.094\pm0.070$ & $-0.036\pm0.127$ & $-0.010\pm0.035$ & $-0.056\pm0.042$ & $-0.086\pm0.028$ \\
  68978A & $-0.058\pm0.017$ & $-0.058\pm0.070$ & $-0.083\pm0.134$ & $ 0.002\pm0.057$ & $-0.003\pm0.007$ & $-0.048\pm0.028$ \\
   69655 & $ 0.026\pm0.021$ & $ 0.143\pm0.070$ & $-0.022\pm0.021$ & $ 0.053\pm0.042$ & $ 0.017\pm0.021$ & $-0.022\pm0.021$ \\
   70889 & $-0.127\pm0.025$ & $-0.044\pm0.070$ & $-0.079\pm0.120$ & $-0.044\pm0.057$ & $-0.064\pm0.014$ & $-0.104\pm0.070$ \\
   71479 & $-0.044\pm0.035$ & $ 0.133\pm0.070$ & $-0.057\pm0.071$ & $ 0.058\pm0.106$ & $-0.027\pm0.057$ & $-0.012\pm0.007$ \\
   73121 & $-0.053\pm0.020$ & $ 0.077\pm0.070$ & $-0.103\pm0.070$ & $ 0.037\pm0.071$ & $ 0.007\pm0.070$ & $ 0.007\pm0.014$ \\
   73524 & $-0.160\pm0.007$ & $-0.055\pm0.070$ & $-0.125\pm0.071$ & $-0.075\pm0.042$ & $-0.065\pm0.085$ & $-0.030\pm0.035$ \\
   88742 & $-0.003\pm0.032$ & $ 0.041\pm0.070$ & $-0.029\pm0.071$ & $-0.029\pm0.057$ & $-0.029\pm0.014$ & $-0.084\pm0.021$ \\
   95456 & $-0.068\pm0.021$ & $-0.012\pm0.070$ & $-0.012\pm0.099$ & $ 0.028\pm0.057$ & $-0.052\pm0.014$ & $-0.082\pm0.028$ \\
    9782 & $-0.002\pm0.045$ & $ 0.015\pm0.070$ & $-0.015\pm0.099$ & $ 0.025\pm0.057$ & $-0.055\pm0.042$ & $-0.060\pm0.035$ \\
   33636 & $-0.106\pm0.015$ & $ 0.061\pm0.070$ & $-0.114\pm0.021$ & $-0.039\pm0.057$ & $ 0.161\pm0.070$ & $-0.054\pm0.021$ \\
\noalign{\medskip} 
\hline
\end{tabular}
\end{table*}

\clearpage

\begin{table*}
\centering
\caption{Abundance ratios [X/Fe] of F-, G-type analogs without known planets\label{tabonline}}
\begin{tabular}{lrrrrrr}
\noalign{\smallskip}
\noalign{\smallskip}
\noalign{\smallskip}
\hline
\hline
\noalign{\medskip} 
HD       &     [Si/Fe]      &     [Ca/Fe]      &     [Sc/Fe]      &     [Ti/Fe]      &     [V/Fe]      &     [Cr/Fe]      \\
\noalign{\medskip} 
\hline
\hline
\noalign{\medskip} 
   11226 & $ 0.012\pm0.011$ & $ 0.003\pm0.017$ & $ 0.032\pm0.051$ & $-0.021\pm0.024$ & $-0.030\pm0.043$ & $-0.007\pm0.017$ \\
  119638 & $ 0.027\pm0.009$ & $ 0.056\pm0.028$ & $ 0.023\pm0.026$ & $ 0.015\pm0.023$ & $-0.042\pm0.026$ & $-0.004\pm0.022$ \\
  122862 & $ 0.027\pm0.009$ & $ 0.041\pm0.031$ & $ 0.071\pm0.026$ & $ 0.015\pm0.013$ & $-0.035\pm0.051$ & $-0.003\pm0.024$ \\
  125881 & $-0.011\pm0.022$ & $ 0.023\pm0.014$ & $-0.019\pm0.037$ & $-0.019\pm0.017$ & $-0.043\pm0.018$ & $ 0.005\pm0.010$ \\
    1388 & $ 0.002\pm0.014$ & $ 0.036\pm0.014$ & $ 0.042\pm0.043$ & $ 0.007\pm0.017$ & $-0.072\pm0.024$ & $-0.007\pm0.020$ \\
  145666 & $-0.031\pm0.010$ & $ 0.019\pm0.013$ & $-0.052\pm0.017$ & $-0.014\pm0.016$ & $-0.077\pm0.030$ & $ 0.008\pm0.012$ \\
  157338 & $-0.002\pm0.009$ & $ 0.024\pm0.023$ & $ 0.009\pm0.024$ & $-0.008\pm0.013$ & $-0.056\pm0.045$ & $-0.006\pm0.022$ \\
    1581 & $ 0.030\pm0.007$ & $ 0.056\pm0.018$ & $ 0.040\pm0.027$ & $ 0.045\pm0.027$ & $-0.033\pm0.055$ & $-0.014\pm0.020$ \\
  168871 & $ 0.017\pm0.017$ & $ 0.056\pm0.034$ & $ 0.057\pm0.043$ & $ 0.007\pm0.016$ & $-0.030\pm0.040$ & $-0.007\pm0.023$ \\
  171990 & $ 0.022\pm0.012$ & $ 0.041\pm0.047$ & $ 0.060\pm0.041$ & $-0.005\pm0.017$ & $-0.028\pm0.029$ & $-0.012\pm0.017$ \\
  193193 & $ 0.010\pm0.009$ & $ 0.015\pm0.018$ & $ 0.063\pm0.038$ & $ 0.005\pm0.016$ & $-0.013\pm0.025$ & $-0.002\pm0.019$ \\
  196800 & $ 0.003\pm0.017$ & $-0.017\pm0.014$ & $ 0.045\pm0.038$ & $-0.025\pm0.030$ & $-0.002\pm0.044$ & $-0.010\pm0.034$ \\
  199960 & $ 0.015\pm0.019$ & $-0.021\pm0.063$ & $ 0.057\pm0.035$ & $-0.017\pm0.035$ & $ 0.007\pm0.032$ & $-0.008\pm0.020$ \\
  204385 & $ 0.011\pm0.017$ & $ 0.012\pm0.022$ & $ 0.042\pm0.027$ & $-0.013\pm0.026$ & $-0.058\pm0.033$ & $ 0.005\pm0.018$ \\
  221356 & $ 0.012\pm0.012$ & $ 0.069\pm0.049$ & $ 0.075\pm0.016$ & $ 0.088\pm0.036$ & $-0.069\pm0.039$ & $-0.025\pm0.031$ \\
   31822 & $-0.001\pm0.014$ & $ 0.054\pm0.025$ & $-0.022\pm0.022$ & $ 0.014\pm0.040$ & $-0.104\pm0.028$ & $ 0.005\pm0.025$ \\
   36379 & $ 0.044\pm0.011$ & $ 0.062\pm0.029$ & $ 0.070\pm0.033$ & $ 0.021\pm0.024$ & $-0.020\pm0.053$ & $ 0.006\pm0.023$ \\
    3823 & $ 0.061\pm0.019$ & $ 0.080\pm0.021$ & $ 0.072\pm0.033$ & $ 0.058\pm0.029$ & $-0.035\pm0.050$ & $-0.008\pm0.020$ \\
   38973 & $ 0.006\pm0.012$ & $ 0.021\pm0.017$ & $ 0.003\pm0.030$ & $-0.023\pm0.019$ & $-0.033\pm0.035$ & $ 0.004\pm0.018$ \\
   44120 & $ 0.004\pm0.019$ & $ 0.035\pm0.049$ & $ 0.031\pm0.022$ & $-0.020\pm0.024$ & $-0.008\pm0.025$ & $-0.020\pm0.020$ \\
   44447 & $ 0.039\pm0.009$ & $ 0.051\pm0.016$ & $ 0.058\pm0.040$ & $ 0.025\pm0.019$ & $-0.033\pm0.055$ & $-0.007\pm0.020$ \\
    6735 & $ 0.011\pm0.012$ & $ 0.057\pm0.021$ & $-0.002\pm0.033$ & $ 0.009\pm0.034$ & $-0.075\pm0.047$ & $-0.010\pm0.019$ \\
  68978A & $ 0.001\pm0.011$ & $ 0.030\pm0.023$ & $-0.003\pm0.017$ & $-0.006\pm0.013$ & $-0.037\pm0.039$ & $ 0.011\pm0.013$ \\
   69655 & $ 0.019\pm0.009$ & $ 0.038\pm0.024$ & $ 0.025\pm0.022$ & $ 0.006\pm0.019$ & $-0.062\pm0.043$ & $-0.004\pm0.013$ \\
   70889 & $-0.028\pm0.008$ & $ 0.021\pm0.024$ & $-0.039\pm0.045$ & $-0.026\pm0.027$ & $-0.064\pm0.023$ & $ 0.008\pm0.014$ \\
   71479 & $-0.001\pm0.019$ & $-0.015\pm0.026$ & $ 0.083\pm0.041$ & $-0.003\pm0.017$ & $ 0.029\pm0.022$ & $-0.000\pm0.019$ \\
   73121 & $ 0.016\pm0.015$ & $ 0.015\pm0.017$ & $ 0.060\pm0.036$ & $-0.000\pm0.024$ & $-0.006\pm0.043$ & $-0.008\pm0.026$ \\
   73524 & $-0.008\pm0.014$ & $ 0.044\pm0.030$ & $ 0.005\pm0.022$ & $ 0.010\pm0.020$ & $-0.030\pm0.024$ & $ 0.006\pm0.016$ \\
   88742 & $-0.014\pm0.013$ & $ 0.019\pm0.026$ & $-0.044\pm0.034$ & $-0.029\pm0.019$ & $-0.043\pm0.023$ & $ 0.013\pm0.016$ \\
   95456 & $ 0.000\pm0.014$ & $ 0.028\pm0.047$ & $-0.009\pm0.015$ & $-0.010\pm0.034$ & $-0.060\pm0.037$ & $-0.003\pm0.024$ \\
    9782 & $-0.014\pm0.020$ & $-0.010\pm0.048$ & $-0.018\pm0.021$ & $-0.045\pm0.018$ & $-0.042\pm0.044$ & $-0.013\pm0.019$ \\
   33636 & $ 0.004\pm0.006$ & $ 0.051\pm0.015$ & $ 0.015\pm0.042$ & $ 0.041\pm0.022$ & $-0.031\pm0.038$ & $ 0.008\pm0.027$ \\
\noalign{\medskip} 
\hline
\end{tabular}
\end{table*}

\clearpage

\begin{table*}
\centering
\caption{Abundance ratios [X/Fe] of F-, G-type analogs without known planets\label{tabonline}}
\begin{tabular}{lrrrrrr}
\noalign{\smallskip}
\noalign{\smallskip}
\noalign{\smallskip}
\hline
\hline
\noalign{\medskip} 
HD       &     [Mn/Fe]      &     [Co/Fe]      &     [Ni/Fe]      &     [Cu/Fe]      &     [Zn/Fe]      &     [Sr/Fe]      \\
\noalign{\medskip} 
\hline
\hline
\noalign{\medskip} 
   11226 & $-0.057\pm0.058$ & $-0.022\pm0.016$ & $ 0.004\pm0.018$ & $ 0.003\pm0.070$ & $ 0.008\pm0.035$ & $-0.036\pm0.070$ \\
  119638 & $-0.097\pm0.036$ & $-0.033\pm0.025$ & $-0.025\pm0.014$ & $ 0.006\pm0.070$ & $-0.014\pm0.042$ & $-0.035\pm0.070$ \\
  122862 & $-0.039\pm0.070$ & $-0.028\pm0.010$ & $-0.010\pm0.013$ & $ 0.044\pm0.070$ & $ 0.019\pm0.007$ & $-0.096\pm0.070$ \\
  125881 & $-0.057\pm0.038$ & $-0.069\pm0.012$ & $-0.024\pm0.015$ & $-0.027\pm0.070$ & $-0.017\pm0.028$ & $ 0.033\pm0.070$ \\
    1388 & $-0.082\pm0.046$ & $-0.038\pm0.012$ & $-0.021\pm0.014$ & $-0.008\pm0.070$ & $-0.043\pm0.021$ & $-0.008\pm0.070$ \\
  145666 & $-0.089\pm0.045$ & $-0.089\pm0.007$ & $-0.048\pm0.015$ & $-0.037\pm0.070$ & $-0.037\pm0.042$ & $ 0.053\pm0.070$ \\
  157338 & $-0.031\pm0.071$ & $-0.054\pm0.024$ & $-0.024\pm0.012$ & $-0.006\pm0.070$ & $ 0.019\pm0.035$ & $-0.026\pm0.070$ \\
    1581 & $-0.118\pm0.039$ & $-0.007\pm0.022$ & $-0.024\pm0.014$ & $-0.003\pm0.070$ & $ 0.007\pm0.014$ & $-0.043\pm0.070$ \\
  168871 & $-0.081\pm0.106$ & $-0.035\pm0.016$ & $-0.012\pm0.013$ & $ 0.027\pm0.070$ & $ 0.037\pm0.028$ & $-0.053\pm0.070$ \\
  171990 & $-0.056\pm0.065$ & $ 0.016\pm0.028$ & $ 0.013\pm0.021$ & $ 0.048\pm0.070$ & $ 0.033\pm0.021$ & $-0.062\pm0.070$ \\
  193193 & $-0.050\pm0.026$ & $-0.016\pm0.021$ & $-0.011\pm0.015$ & $ 0.005\pm0.070$ & $ 0.015\pm0.014$ & $-0.065\pm0.070$ \\
  196800 & $ 0.015\pm0.015$ & $ 0.015\pm0.019$ & $ 0.028\pm0.017$ & $ 0.013\pm0.070$ & $ 0.013\pm0.057$ & $-0.057\pm0.070$ \\
  199960 & $ 0.087\pm0.025$ & $ 0.059\pm0.013$ & $ 0.056\pm0.028$ & $ 0.059\pm0.070$ & $ 0.069\pm0.113$ & $-0.091\pm0.070$ \\
  204385 & $-0.024\pm0.048$ & $ 0.009\pm0.014$ & $ 0.001\pm0.020$ & $ 0.032\pm0.070$ & $ 0.036\pm0.092$ & $-0.128\pm0.070$ \\
  221356 & $-0.155\pm0.073$ & $-0.035\pm0.035$ & $-0.020\pm0.037$ & $ 0.075\pm0.070$ & $-0.105\pm0.057$ & $ 0.005\pm0.070$ \\
   31822 & $-0.162\pm0.054$ & $-0.086\pm0.033$ & $-0.057\pm0.015$ & $-0.004\pm0.070$ & $-0.099\pm0.049$ & $-0.004\pm0.070$ \\
   36379 & $-0.069\pm0.078$ & $-0.020\pm0.023$ & $-0.015\pm0.015$ & $ 0.030\pm0.070$ & $ 0.000\pm0.070$ & $-0.100\pm0.070$ \\
    3823 & $-0.153\pm0.027$ & $-0.002\pm0.016$ & $-0.013\pm0.015$ & $ 0.067\pm0.070$ & $-0.028\pm0.078$ & $-0.093\pm0.070$ \\
   38973 & $-0.045\pm0.037$ & $-0.031\pm0.013$ & $-0.001\pm0.013$ & $-0.005\pm0.070$ & $ 0.010\pm0.035$ & $-0.055\pm0.070$ \\
   44120 & $-0.016\pm0.055$ & $-0.016\pm0.008$ & $ 0.013\pm0.015$ & $ 0.014\pm0.070$ & $-0.001\pm0.078$ & $-0.076\pm0.070$ \\
   44447 & $-0.120\pm0.039$ & $-0.029\pm0.033$ & $-0.025\pm0.015$ & $ 0.020\pm0.070$ & $ 0.015\pm0.021$ & $-0.130\pm0.070$ \\
    6735 & $-0.144\pm0.084$ & $-0.061\pm0.032$ & $-0.042\pm0.016$ & $-0.025\pm0.070$ & $-0.050\pm0.007$ & $-0.005\pm0.070$ \\
  68978A & $-0.082\pm0.044$ & $-0.028\pm0.019$ & $-0.017\pm0.014$ & $-0.028\pm0.070$ & $-0.143\pm0.007$ & $-0.218\pm0.070$ \\
   69655 & $-0.095\pm0.039$ & $-0.011\pm0.015$ & $-0.016\pm0.018$ & $ 0.013\pm0.070$ & $ 0.032\pm0.085$ & $-0.057\pm0.070$ \\
   70889 & $-0.024\pm0.024$ & $-0.082\pm0.016$ & $-0.033\pm0.009$ & $-0.064\pm0.070$ & $-0.074\pm0.028$ & $ 0.046\pm0.070$ \\
   71479 & $ 0.037\pm0.043$ & $ 0.033\pm0.018$ & $ 0.041\pm0.015$ & $ 0.023\pm0.070$ & $ 0.003\pm0.042$ & $-0.037\pm0.070$ \\
   73121 & $-0.033\pm0.033$ & $-0.021\pm0.013$ & $ 0.001\pm0.017$ & $ 0.007\pm0.070$ & $ 0.002\pm0.035$ & $-0.033\pm0.070$ \\
   73524 & $-0.060\pm0.024$ & $-0.022\pm0.013$ & $-0.012\pm0.015$ & $-0.045\pm0.070$ & $-0.095\pm0.042$ & $ 0.055\pm0.070$ \\
   88742 & $-0.057\pm0.022$ & $-0.071\pm0.022$ & $-0.034\pm0.014$ & $-0.069\pm0.070$ & $-0.044\pm0.007$ & $ 0.031\pm0.070$ \\
   95456 & $-0.029\pm0.039$ & $-0.035\pm0.022$ & $-0.010\pm0.018$ & $-0.002\pm0.070$ & $-0.047\pm0.007$ & $-0.012\pm0.070$ \\
    9782 & $-0.020\pm0.029$ & $-0.038\pm0.020$ & $-0.005\pm0.019$ & $-0.025\pm0.070$ & $-0.050\pm0.035$ & $-0.025\pm0.070$ \\
   33636 &        --        & $-0.062\pm0.016$ & $-0.047\pm0.017$ & $-0.079\pm0.070$ & $-0.039\pm0.070$ &        --        \\
\noalign{\medskip} 
\hline
\end{tabular}
\end{table*}

\clearpage

\begin{table*}
\centering
\caption{Abundance ratios [X/Fe] of F-, G-type analogs without known planets\label{tabonline}}
\begin{tabular}{lrrrrrr}
\noalign{\smallskip}
\noalign{\smallskip}
\noalign{\smallskip}
\hline
\hline
\noalign{\medskip} 
HD       &     [Y/Fe]      &     [Zr/Fe]      &     [Ba/Fe]      &     [Ce/Fe]      &     [Nd/Fe]      &     [Eu/Fe]      \\
\noalign{\medskip} 
\hline
\hline
\noalign{\medskip} 
   11226 & $ 0.030\pm0.029$ & $ 0.044\pm0.070$ & $ 0.048\pm0.106$ & $-0.047\pm0.030$ & $-0.117\pm0.070$ & $ 0.003\pm0.070$ \\
  119638 & $ 0.056\pm0.035$ & $ 0.096\pm0.070$ & $ 0.126\pm0.042$ & $-0.001\pm0.076$ & $-0.054\pm0.070$ & $ 0.046\pm0.070$ \\
  122862 & $ 0.000\pm0.046$ & $ 0.034\pm0.070$ & $ 0.024\pm0.028$ & $-0.046\pm0.060$ & $-0.056\pm0.070$ & $ 0.074\pm0.070$ \\
  125881 & $ 0.103\pm0.010$ & $ 0.133\pm0.070$ & $ 0.148\pm0.049$ & $ 0.019\pm0.038$ & $-0.027\pm0.070$ & $ 0.063\pm0.070$ \\
    1388 & $ 0.048\pm0.021$ & $ 0.072\pm0.070$ & $ 0.067\pm0.007$ & $ 0.028\pm0.025$ & $-0.078\pm0.070$ & $ 0.042\pm0.070$ \\
  145666 & $ 0.119\pm0.012$ & $ 0.133\pm0.070$ & $ 0.198\pm0.007$ & $ 0.096\pm0.029$ & $ 0.003\pm0.070$ & $ 0.033\pm0.070$ \\
  157338 & $ 0.057\pm0.015$ & $ 0.074\pm0.070$ & $ 0.129\pm0.106$ & $-0.003\pm0.025$ & $-0.086\pm0.070$ & $ 0.014\pm0.070$ \\
    1581 & $ 0.040\pm0.023$ & $ 0.107\pm0.070$ & $ 0.097\pm0.028$ & $ 0.040\pm0.050$ & $ 0.007\pm0.070$ & $ 0.097\pm0.070$ \\
  168871 & $ 0.024\pm0.029$ & $ 0.057\pm0.070$ & $ 0.052\pm0.021$ & $-0.006\pm0.021$ & $-0.083\pm0.070$ & $ 0.077\pm0.070$ \\
  171990 & $ 0.044\pm0.038$ & $ 0.048\pm0.070$ & $ 0.043\pm0.021$ & $-0.032\pm0.061$ & $-0.102\pm0.070$ & $ 0.028\pm0.070$ \\
  193193 & $ 0.015\pm0.026$ & $-0.015\pm0.070$ & $ 0.060\pm0.021$ & $-0.018\pm0.029$ & $-0.095\pm0.070$ & $-0.015\pm0.070$ \\
  196800 & $ 0.013\pm0.026$ & $-0.067\pm0.070$ & $-0.037\pm0.014$ & $-0.034\pm0.092$ & $-0.127\pm0.070$ & $-0.167\pm0.070$ \\
  199960 & $-0.008\pm0.075$ & $ 0.009\pm0.070$ & $-0.076\pm0.035$ &        --        & $-0.161\pm0.070$ & $ 0.059\pm0.070$ \\
  204385 & $ 0.045\pm0.023$ & $-0.048\pm0.070$ & $ 0.061\pm0.028$ & $-0.035\pm0.050$ & $-0.088\pm0.070$ & $-0.008\pm0.070$ \\
  221356 & $ 0.052\pm0.061$ & $ 0.075\pm0.070$ & $ 0.215\pm0.014$ &        --        & $ 0.065\pm0.070$ & $ 0.145\pm0.070$ \\
   31822 & $ 0.092\pm0.012$ & $ 0.096\pm0.070$ & $ 0.271\pm0.035$ & $ 0.122\pm0.086$ & $ 0.066\pm0.070$ & $-0.094\pm0.070$ \\
   36379 & $-0.020\pm0.044$ & $ 0.020\pm0.070$ & $ 0.030\pm0.042$ & $-0.050\pm0.026$ & $-0.089\pm0.070$ & $ 0.041\pm0.070$ \\
    3823 & $-0.043\pm0.053$ & $ 0.017\pm0.070$ & $-0.028\pm0.021$ & $-0.033\pm0.050$ & $-0.083\pm0.070$ & $ 0.077\pm0.070$ \\
   38973 & $ 0.052\pm0.021$ & $ 0.035\pm0.070$ & $ 0.030\pm0.035$ & $-0.042\pm0.035$ & $-0.075\pm0.070$ & $ 0.025\pm0.070$ \\
   44120 & $ 0.014\pm0.026$ & $ 0.014\pm0.070$ & $ 0.019\pm0.007$ & $-0.046\pm0.020$ & $-0.116\pm0.070$ & $ 0.034\pm0.070$ \\
   44447 & $-0.047\pm0.049$ & $-0.010\pm0.070$ & $-0.010\pm0.014$ & $-0.050\pm0.062$ & $-0.100\pm0.070$ & $ 0.030\pm0.070$ \\
    6735 & $ 0.101\pm0.029$ & $ 0.134\pm0.070$ & $ 0.194\pm0.028$ & $ 0.041\pm0.137$ & $-0.036\pm0.070$ & $-0.005\pm0.070$ \\
  68978A & $ 0.079\pm0.029$ & $ 0.042\pm0.070$ & $ 0.122\pm0.028$ & $ 0.032\pm0.017$ & $-0.068\pm0.070$ & $-0.098\pm0.070$ \\
   69655 & $ 0.002\pm0.026$ & $ 0.023\pm0.070$ & $ 0.058\pm0.007$ & $-0.007\pm0.017$ & $-0.047\pm0.070$ & $ 0.032\pm0.070$ \\
   70889 & $ 0.113\pm0.012$ & $ 0.116\pm0.070$ & $ 0.176\pm0.028$ & $ 0.053\pm0.015$ & $-0.084\pm0.070$ & $-0.054\pm0.070$ \\
   71479 & $ 0.010\pm0.051$ & $-0.017\pm0.070$ & $-0.047\pm0.028$ & $-0.044\pm0.049$ & $-0.157\pm0.070$ & $-0.027\pm0.070$ \\
   73121 & $ 0.067\pm0.044$ & $ 0.047\pm0.070$ & $ 0.107\pm0.042$ & $ 0.020\pm0.046$ & $-0.083\pm0.070$ & $ 0.067\pm0.070$ \\
   73524 & $ 0.085\pm0.040$ & $ 0.105\pm0.070$ & $ 0.095\pm0.014$ & $ 0.042\pm0.025$ & $-0.055\pm0.070$ & $ 0.045\pm0.070$ \\
   88742 & $ 0.097\pm0.021$ & $ 0.131\pm0.070$ & $ 0.151\pm0.014$ & $ 0.051\pm0.030$ & $-0.049\pm0.070$ & $-0.039\pm0.070$ \\
   95456 & $ 0.078\pm0.044$ & $ 0.058\pm0.070$ & $ 0.028\pm0.070$ & $-0.025\pm0.032$ & $-0.052\pm0.070$ & $-0.052\pm0.070$ \\
    9782 & $ 0.058\pm0.015$ & $ 0.045\pm0.070$ & $ 0.030\pm0.021$ & $-0.065\pm0.078$ & $-0.145\pm0.070$ & $-0.045\pm0.070$ \\
   33636 & $ 0.101\pm0.017$ & $ 0.141\pm0.070$ & $ 0.191\pm0.014$ &        --        & $ 0.071\pm0.070$ & $ 0.051\pm0.070$ \\
\noalign{\medskip} 
\hline
\end{tabular}
\end{table*}

\clearpage

\end{appendix}

\end{document}